\documentclass[preprint,12pt,authoryear]{elsarticle}
\usepackage[english]{babel}
\usepackage{caption}
\usepackage{subcaption}
\usepackage[a4paper,top=2cm,bottom=2cm,left=3cm,right=3cm,marginparwidth=1.75cm]{geometry}

\usepackage{amssymb}
\usepackage{amsmath}
\usepackage{graphicx}
\usepackage[colorlinks=true, allcolors=blue]{hyperref}
\usepackage{color}
\usepackage{soul}
\usepackage{wasysym}
\usepackage{verbatim} 

\usepackage[percent]{overpic}

\biboptions{authoryear,round}

\journal{Journal of Computational Physics}

\begin{document}

\begin{frontmatter}

\title{A geometric interpolation scheme for applying dynamic wetting to three-dimensional volume of fluid simulations}

\author[1]{Yifan Han}

\author[1]{Gerd Mutschke\corref{cor1}}
\author[1,2,3]{Kerstin Eckert}

\cortext[cor1]{Corresponding author: \href{mailto:g.mutschke@hzdr.de}{g.mutschke@hzdr.de}}

\affiliation[1]{organization={Institute of Fluid Dynamics, Helmholtz-Zentrum Dresden-Rossendorf},%Department and Organization
            addressline={Bautzner Landstrasse 400}, 
            city={Dresden},
            postcode={01328}, 
            country={Germany}}

\affiliation[2]{organization={Institute of Process Engineering and Environmental Technology, Technische Universit\"at Dresden},%Department and Organization
            %addressline={Bautzner Landstrasse 400}, 
            city={Dresden},
            postcode={01062}, 
            country={Germany}}

\affiliation[3]{organization={Hydrogen Lab, School of Engineering, Technische Universit\"at Dresden},%Department and Organization
            %addressline={Bautzner Landstrasse 400}, 
            city={Dresden},
            postcode={01062}, 
            country={Germany}}

\begin{abstract}
This paper presents a three-dimensional framework for simulating dynamic wetting phenomena using the volume of fluid (VOF) method, implemented in Basilisk. 
A geometric interpolation scheme is developed 
to obtain an accurate and reliable
value of the contact line velocity.
To capture realistic wetting dynamics, a dynamic contact angle model is integrated
that considers also
contact angle hysteresis (CAH).
The approach is validated against 
various experimental results, including droplet spreading, splashing, and sliding 
and demonstrates quantitative agreement with the three-dimensional wetting behavior observed.
Additionally, a comparative analysis between dynamic and static contact angle models is performed.
%\textcolor{blue}{
%Above all, the simulation results reveal that the proposed interpolation scheme significantly improves numerical accuracy in contact line velocity estimation while enabling the use of larger time steps.}
%\textcolor{red}{(Gerd: Do we show this somewhere???)}
%Finally, the study explores potential applications such as bubble coalescence on a surfaces and bubble detachment under shear flow with dynamic wetting.

\end{abstract}

%% Keywords
%{\it Keywords: dynamic wetting, contact line velocity, geometric interpolation, VOF}

\end{frontmatter}

\section{Introduction}
\label{Introduction}

Contact line dynamics, describing the motion of the intersection between gas, liquid, and solid phases, plays a pivotal role in a wide range of multiphase flow phenomena.
In the context of bubbles and droplets, 
it
critically influences interfacial deformation
and interaction with solid substrates.
At the heart of these processes lies dynamic wetting, where the contact angle evolves in response to local flow conditions such as viscous dissipation, capillary force, and surface adhesion. 
This dynamic behavior
becomes especially important when 
the local conditions are strongly changing,
e.g.~during droplet spreading and splashing or bubble growth at laterally inhomogeneous surfaces \citep{luo2023re,de2021droplet,quetzeri2019role,heinrich2024functionalization}.

To accurately describe
dynamic wetting, a variety of modeling strategies have been developed across different length scales. At the microscopic scale, 
the Molecular Kinetic Theory (MKT) describes contact line motion in terms of the statistical mechanics of molecular displacements on the solid surface~\citep{hayes1994molecular,blake2006physics}.
Building on this theory, the Generalized Navier Boundary Condition (GNBC) model \citep{zhang2017multiscale,liu2021multiscale,kulkarni2023stream,ESTEBAN2023105946,fullana2024consistent} has been introduced for continuum-scale simulations. 
It couples the wall slip velocity to both the viscous stress and the imbalance between dynamic and static contact angle. 
This makes it particularly suitable for molecular to mesoscopic flow scales, where near-wall dynamics, interfacial slip, and contact line forces play a dominant role.
At the macroscopic scale, the classical Cox-Voinov model~\citep{voinov1976hydrodynamics,cox1986dynamics} describes how the apparent contact angle depends on the contact line velocity, accounting for viscous bending of the fluid interface near the wall. 
This theoretical insight has motivated the development of dynamic contact angle models \citep{popescu2008capillary,seveno2009dynamics,dwivedi2022dynamic}, based on the original Cox-Voinov model.
In general, the GNBC model is most appropriate for problems where molecular-level wetting dynamics and interfacial slip are dominant—such as nanoscale flows or wetting on chemically heterogeneous surfaces. 
In contrast, the velocity-based dynamic contact angle models are more suitable for macroscopic applications involving large interface deformation, such as droplet impact and bubble detachment. 
The latter models are easily implemented into numerical methods with explicit characterization of contact line position and velocity, including the level-set method \citep{SPELT2005389,yokoi2009numerical,Park2012Num,Xu_Ren_2016,ZHANG2020109636}, diffuse interface methods~\citep{ding2007wetting,yue2010sharp,yue2011wall}, and finite element methods \citep{dwivedi2022dynamic,qin2024numerical}.

The Volume-of-Fluid (VOF) method is a robust and widely adopted approach for simulating multiphase flows with sharp fluid interfaces. 
Its strength lies in handling complex interface dynamics, including topological changes such as breakup, coalescence, and spreading. 
In the open-source solver Basilisk~\citep{POPINET20095838}, 
the VOF method is implemented together with adaptive mesh refinement (AMR), allowing for high-resolution tracking of interfacial features and steep near-wall gradients.
This computational framework has been extensively applied to a wide range of problems, including  mass transfer across interfaces~\citep{cipriano2024multicomponent,long2024edge,xue2023three,gennari2022phase,farsoiya2023direct}, and fluid–solid interactions such as wetting.
Early implementations focused on static contact angles imposed via height-function method, enabling the simulation of equilibrium wetting states on smooth surfaces~\citep{afkhami2008height,AfkhamiBussmann2009,han2021consistent}.
More recently, the height-function method has been extended to handle wetting on structured or rough surfaces through the integration of embedded boundary techniques~\citep{tavares2024coupled,huang20252d,chen2025volume}.
While these developments have significantly improved the modeling of static wetting on both smooth and structured surfaces, 
realistic interfacial dynamics under non-equilibrium conditions demand a dynamic contact angle model.

To accurately model dynamic wetting behavior in the VOF framework, 
a critical requirement is the precise determination of the local contact line velocity, which governs the evolution of the apparent contact angle.
Afkhami et al.~\citep{afkhami2009mesh} proposed a mesh-dependent dynamic wetting model, where the contact line velocity is defined as the tangential fluid velocity at the center of the first grid cell adjacent to the wall at the interface.
This model was later extended to investigate the transition from wetting to forced dewetting 
according to the Cox model \citep{afkhami2018transition}.
Several researchers have proposed refined strategies for estimating the contact line velocity in algebraic VOF-based solvers.
In three-dimensional configurations, the contact line velocity is computed by projecting the interface cell velocity onto the local interface normal direction~\citep{malgarinos2014vof,linder20153d,gohl2018immersed}.
However, in the sharp-interface VOF framework, the interface is not explicitly located at 
$f$ =0.5, making it challenging to accurately determine the velocity at the interface.
To address this challenge, Dupont and Legendre \citep{DUPONT20102453} estimated the contact line velocity in a 2D setup by interpolating the fluid velocity at the iso-contour corresponding to a volume fraction of $f$ =0.5 in the code JADIM.
Fullana et al.~\citep{fullana2025mass} recently implemented a toy model in Basilisk for dynamic wetting, where the contact line velocity is estimated as the tangential displacement of the interface along a virtual boundary over successive time steps.
In our previous work, the 
2D
contact line velocity was
approximated 
by averaging
the center velocities of the neighboring
interface cells at the wall
\citep{han2025numerical}.
Huang et al.~\citep{huang20252d} extended this approximation in 2D
to  surfaces that may be inclined,
thus intersecting
the mesh in Basilisk.
%: they first compute an interface-averaged fluid velocity from interfacial cells adjacent to the contact-line cell that do not intersect the solid, and then obtain the surface-parallel contact-line velocity by projecting this average onto the wall-tangent direction.
%
%Huang et al.~\citep{huang20252d} extended this approximation to take into account in 2D also inclined surfaces, where the surface-parallel contact line velocity is computed by averaging the tangential component of the fluid velocity in those interfacial cells without solid boundary and neighboring the contact line cell.
%over the adjacent interfacial cell without solid fraction.
However, the existing approaches are largely restricted to two-dimensional or simplified three-dimensional configurations, and a general, robust method for accurately determining the contact line velocity in three-dimensional sharp-interface VOF models is lacking.

%To overcome these limitations, 
The present work introduces a geometric interpolation scheme for estimating the contact line velocity in such three-dimensional problems, implemented in the Basilisk framework. 
A dynamic contact angle (DCA) model incorporating contact angle hysteresis (CAH) is integrated to provide physically sound boundary conditions at the contact line. 
The proposed method is designed to improve both the accuracy and robustness of dynamic wetting simulations, enabling the use of coarser meshes and larger time steps without sacrificing fidelity.
The remainder of this paper is organized as follows.  
Section \ref{Methodology} describes in detail the numerical methodology and its implementation in Basilisk. 
Section~\ref{drop-impact} examines droplet impact on solid substrates, encompassing the spreading and splashing regimes. 
Section \ref{drop-slide-NC2023} investigates gravity-driven droplet sliding on inclined surfaces. 
Finally, Section \ref{conclusion} summarizes the findings and discusses potential directions for future work.

\section{Numerical methodology}
\label{Methodology}

\subsection{Governing equations}
\label{NSeqs}
%================================
In the Volume of Fluid (VOF) method, a scalar volume fraction 
$f$ is defined to distinguish between fluid phases,
such that $f = 0 $  represents the gas phase and 
$f = 1 $ corresponds to the liquid phase.
The subsequent analysis employs the one-fluid formulation, 
where density $\rho$, velocity $\vec{u}$, pressure $p$, and viscosity $\mu$ are expressed as weighted averages of the corresponding properties in the gas and liquid phases, with the weighting factor being the volume fraction $f$ \citep{kataoka1986}. 
Specifically, the density and viscosity fields are given by:
%-----------------------------------------
\begin{equation}
    \rho = (1-f) \rho_g + f \rho_l, \quad \mu = (1-f) \mu_g + f \mu_l. 
   \label{eq-vof}
\end{equation}
%-----------------------------------------
%\noindent
%-----------------------------------------
%\begin{equation}
%    \vec{u} = (1-f) \vec{u}_g + f \vec{u}_l 
%    \label{eq04a}
%\end{equation}
%-----------------------------------------
where the subscripts $g$ and $l$ denote the gas and the 
liquid phase, respectively.
%Introducing the one-fluid formulation
This formulation enables the mass conservation and Navier–Stokes equations across the gas–liquid interface $\Sigma$ to be expressed as follows: %\citep{gennari2022phase}
%-----------------------------------------
\begin{align}
      & \nabla  \cdot  \vec{u} =   0, \label{eq02} \\
      &  \frac{\partial \vec{u} }{\partial t} +  \vec{u} \cdot  \nabla \vec{u}    = -\frac{1}{\rho}\nabla p + \frac{1}{\rho}\nabla \cdot \left \{    \mu [ \nabla \vec{u} +(\nabla \vec{u})^T ] \right \}  +  \vec{g} + \frac{\gamma \kappa {\vec{n}_{\Sigma}}}{\rho} \delta_{\Sigma},
    \label{eq-NS-eq}
\end{align}
%-----------------------------------------
\noindent
where $\vec{g}$ denotes the gravitational acceleration vector.
$\gamma$ and $\kappa$ denote the surface tension coefficient and curvature of the gas-liquid interface, and $\vec{n}$ is the unit normal vector oriented perpendicular to the interface, pointing from the liquid towards the gas phase.
The term $\delta_{\Sigma}$ represents the Dirac delta function, 
which is nonzero exclusively at the interface.

The transport equation of the volume fraction $f$ is given as
%-----------------------------------------
\begin{equation}
    \frac{\partial f}{\partial t}  +\vec{u} \cdot \nabla f = 0. 
    \label{eq-phase-fraction} 
\end{equation}
%-----------------------------------------
\noindent
The interface is reconstructed geometrically, 
based on the piecewise linear interface construction (PLIC) method. 
The surface tension force is evaluated according to
the classical continuum surface force (CSF) model \citep{brackbill1992,popinet2018numerical}:
\begin{equation}
    \gamma  \kappa {\vec{n}} \delta_{\Sigma} = \gamma  \kappa \nabla f.
    \label{eq-fsigma}
\end{equation}

%---------------------------------------------------------%
%---------------------------------------------------------%
%---------------------------------------------------------%
\subsection{Geometric interpolation for contact line velocity}
\label{interpol-Ucl}

While the PLIC method provides a consistent framework for reconstructing the
interface location,
the accurate velocity estimation 
at the moving triple phase boundary remains a critical challenge.
In the VOF framework,
the contact line velocity must be resolved at the grid scale to avoid spurious currents.
A common approach is to extract the velocity from the center grid point adjacent to the wall boundary~\citep{afkhami2009mesh,malgarinos2014vof,zhang2017multiscale,gohl2018immersed},
avoiding singularities even under a no-slip boundary condition.
However, due to the 
discontinuity of the velocity field across the interface, 
the cell-averaged velocity in the heavier fluid must be extrapolated to the interface for accurate contact line velocity estimation~\citep{roisman2008drop}.

To address this, we develop a geometric interpolation scheme that systematically reconstructs the velocity field at the interface, ensuring a more precise evaluation of the contact line velocity.
This reconstruction is critical because the dynamic contact angle $\theta_{app}$ depends fundamentally on the contact line motion through the relationship:
\begin{equation}
    \theta_{app} = \theta(\mathrm{Ca}).
    \label{eq-theta_app}
\end{equation}
\noindent
Here, the non-dimensional
capillary number $\mathrm{Ca} =\mu u_{cl}/ \gamma$ 
is determined by
the contact line velocity $u_{cl}$
beside quantifying
the relative importance of viscous to surface tension forces. 
This dependence originates from hydrodynamic bending of the interface (Cox-Voinov theory)~\citep{voinov1976hydrodynamics,cox1986dynamics} and molecular kinetics~\citep{hayes1994molecular,blake2006physics} at the contact line.
In later sections, we will validate the numerical approach by using a specific theoretical model that 
can represent the wettability of
different surfaces.

\begin{figure}
         \centering
         \begin{subfigure}[b]{0.45\textwidth}
             \centering
             \includegraphics[width=\textwidth]{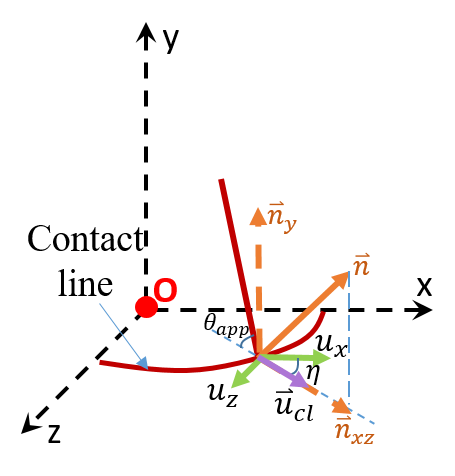}
             \caption{}
             \label{fig-3d-CL-sketch}
         \end{subfigure}
         \hfill
         \begin{subfigure}[b]{0.45\textwidth}
             \centering
             \includegraphics[width=\textwidth]{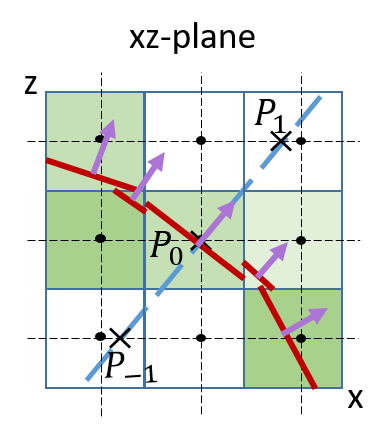}
             \caption{}
             \label{fig-xzplane-sketch}
         \end{subfigure}
            \caption{(a) 3D schematic of the contact line and velocity decomposition. (b) Schematic of interface reconstruction and contact line velocity in the $xz$-plane.
            The green cells represent the interfacial cells.
            %\textcolor{red}{(If the intensity of the green color represents f, then it should be dark green on one side.)}
            }
    \end{figure}

In the following, we consider planar horizontal surfaces.
Figure~\ref{fig-3d-CL-sketch} presents a schematic illustration of the three-phase contact line (red curve) in a three-dimensional coordinate system ($x, y, z$), where
$y$ is the vertical axis
antiparallel to
gravity, and the
$x$ and $z$ axes
span the surface plane.
The contact line is marked by the red curve, and
at the contact point considered,
the red slanted line represents the local tangent to the gas-liquid interface.
$\vec{n}$ corresponds to the local interface normal, and 
$\vec{n}_{xz}$, $\vec{n}_{y}$
denote the projections of 
$\vec{n}$ onto the 
 $xz$-plane and the $y$ axis, respectively.
These projection vectors are not unit vectors in general.
The interface-normal unit vector
$\vec{e}$ in the $x-z$ plane can be obtained
from the $x$ and $z$ components
$n_x$ and $n_z$ of the vector
$\vec{n}_{xz}$. When denoting
the angle between the x-axis
and $\vec{n}_{xz}$ by $\eta$,
it follows
\begin{equation}
     \eta = \mathrm{atan} \left( n_z/n_x \right).
    \label{eq05}
\end{equation}
The unit vector $\vec{e}$ can
then be written as
\begin{equation}
\vec{e} =  
\frac{\vec{n}_{xz}}{|\vec{n}_{xz}|} =
(\cos{\eta}, \sin{\eta})
\end{equation}
%The unit vector normal to the
%interface in the x-z plane
The contact line velocity $\vec{u}_{cl}$ is 
parallel to 
%$\vec{n}_{xz}$  
$\vec{e}$
and is finally
obtained by projecting the 
velocity vector at the interface
$\vec{u}_{\Sigma}$
onto the interface-normal direction $\vec{e}$ in the surface plane.
The components of this velocity
vector at the interface will be
determined by interpolation below.
As the contact line moves 
in the $x-z$-plane,
only the $u_x$ and $u_z$
components are needed, and the projection can be expressed as
%\st{This projection ensures that $\vec{u}_{cl}$ captures the normal motion of the contact line along the interface, written as:}
\begin{equation}
    \vec{u}_{cl} =  
    (\vec{u}_{\Sigma}\cdot\vec{e})\,
    \vec{e} =
    u_{cl} \, \vec{e};
    \qquad u_{cl} =
    u_x \cos{\eta} + u_z \sin{\eta}.
    \label{eq06}
\end{equation}
\noindent
%\st{
%where $\eta$ is the orientation angle of the normal vector $\vec{n}_{xz}$ at the contact point, defined as
%\begin{equation}
%     \eta = \mathrm{atan} \left( %n_z/n_x \right).
%    \label{eq05old}
%\end{equation}
%Here, the unit vector $\vec{e}$ %is given by
%$\vec{e} =  (\cos{\eta}, %\sin{\eta})$.
%}
Finally, the apparent contact angle $\theta_{app}$ 
is defined as the angle
between 
$-\vec{n}_{xz}$ and the red
tangent line at the interface.
%\st{the interface's tangent and the contact line's horizontal normal.}

%--------------------------------------------
%--------------------------------------------

\begin{figure}
         \centering
         \begin{subfigure}[b]{0.4\textwidth}
             \centering
             \includegraphics[width=\textwidth]{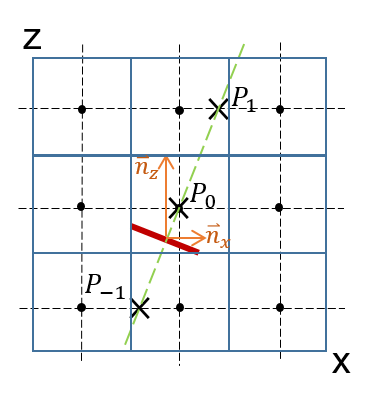}
             \caption{}
             \label{fig-interpo-nz-a}
         \end{subfigure}
         \hfill
         \begin{subfigure}[b]{0.4\textwidth}
             \centering
             \includegraphics[width=\textwidth]{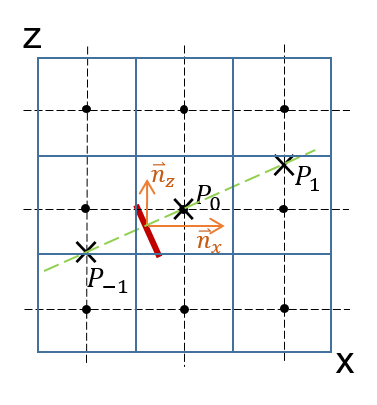}
             \caption{}
             \label{fig-interpo-nx-b}
         \end{subfigure}
            \caption{Interface interpolation for the contact line velocity in the xz-plane: (a) normal along the x-direction; (b) normal along the z-direction.}
            \label{fig-interpo-x-z}
    \end{figure}

Drawing on the unsplit geometrical approaches for concentration gradients at the interface~\citep{bothe2013volume,gennari2022phase},
we adopt a linear geometrical interpolation method for the contact line velocity.
Figure~\ref{fig-xzplane-sketch} depicts a two-dimensional cross-section in the $x-z$-plane.
The computational grid is shaded in varying intensities of green to represent the volume fraction $f$, 
with intermediate values indicating the presence of an interface. 
A red line, obtained via the PLIC method, precisely reconstructs the interface position between gas and liquid phases. 
Purple arrows indicate the contact line velocity vectors $\vec{u}_{cl}$ at the interface.
The three points $P_0$, $P_{1}$, and $P_{-1}$ represent specific locations along the interfacial normal in the computational domain.
$P_0$ is the center point of the grid cell ($i,j,k$).
It serves as the reference point for later determining the local interface orientation and velocity.
Unlike the true interface normal $\vec{n}_{xz}$, 
which can have both positive and negative orientations, 
$P_{1}$, and $P_{-1}$ are always defined as the forward and backward points relative to $P_0$, akin to a standard finite-difference stencil in a structured grid.
The positions
of these two points are 
obtained by
interpolation using
neighboring cells 
and allow
to accurately determine
the velocity near the interface.
The algorithm for computing 
the
contact line velocity via the geometrical interpolation method can be generalized as follows:
\begin{enumerate}

\item Identify all the interfacial cells ($f\in (0,1)$) at the first grid layer adjacent to the x-z plane.  

\item Reconstruct the interface from the volume fraction field $f$ to compute the interface normal vector $\vec{n}$ of each cell.

\item For each interfacial cell, compare the absolute values of $n_x$ and $n_z$ to decide whether to interpolate the values ($f$, $u_x$, $u_z$) at $P_{-1}$ and $P_{1}$ along the $x$ or $z$ direction.
    \begin{itemize}
        \item If $|n_x|<|n_z|$,
        then $P_{-1}$ and $P_{1}$
        are defined at the positions where the
        interface normal
        crosses the horizontal lines in $x$-direction connecting the cell centers below and above the interfacial cell,
         (see Fig.~\ref{fig-interpo-nz-a}).

        \item otherwise,
        $P_{-1}$ and $P_{1}$ are
        defined 
        along the $z$ direction,
        where the interface normal crosses
        the vertical lines connecting the cell centers
        left and right
        (Fig.~\ref{fig-interpo-nx-b}).
    \end{itemize}

\item Compute the gradient of $f$ in 
the normal direction of each cell at the interface by using a central difference approximation:
\begin{equation}
   \frac{\partial f}{\partial n}    \approx \frac{f(P_{1}) - f(P_{-1})}{\overline{P_{-1} P_{1}}}. \quad
  % \frac{\partial f}{\partial z}    \approx \frac{f[0,0,1] - f[0,0,-1]}{2\overline{P_{\1}P_{-1} }.
  \label{eq-dfdx}
\end{equation}
\noindent
where $\overline{P_{-1} P_{1}}$ represents the distance of between the two interpolated points.

\item Perform a linear interpolation of the $x$ and $z$ velocity components, assuming the interface is located at $f=0.5$.
\begin{align}
   & u_{x/z,\mathrm{interp}} = u_{x/z}(P_0) + \frac{u_{x/z}(P_0) - u_{x/z}(P_{\mathrm{adjacent}})}{f(P_0) - f(P_{\mathrm{adjacent}})}(0.5-f(P_0)), \label{eq-ux-interp} 
\end{align}
\noindent
where the point of the adjacent cell $P_{\mathrm{adjacent}}$ 
is selected based on the liquid-phase velocity interpolation direction, determined by the gradient $\frac{\partial f}{\partial n}$:
    \begin{itemize}
    \item Backward interpolation (using $P_{-1}$) is applied if $\frac{\partial f}{\partial n} <0$;
    \item Forward interpolation (using $P_{1}$) is used if $\frac{\partial f}{\partial n}>0$.
   % \item Backward interpolation (using $P_{-1}$) is applied if $f(P_0)>0.5$ and $\frac{\partial f}{\partial n} <0$ or $f(P_0)\le 0.5$ and $\frac{\partial f}{\partial n}>0$;
   % \item Forward interpolation (using $P_{1}$) is used otherwise.
    \end{itemize}
    
\item Compute the contact line velocity by projecting the interpolated velocity components, $u_{x,\mathrm{interp}}$ and $u_{z,\mathrm{interp}}$, onto the interface normal direction via Eq.~\ref{eq06}.

\end{enumerate}

%--------------------------------------------
%--------------------------------------------

In summary, this approach leverages the PLIC-reconstructed interface topology to: (i) determine optimal interpolation directions through normal vector analysis ($|n_x|$ vs. $|n_z|$), (ii) implement directional velocity interpolation using strategically positioned points ($P_{-1}$, $P_{0}$, $P_{1}$), and (iii) project the refined velocities onto the contact line normal direction.

%%%%%%%%%%%%%%%%%%%---------------------------------%%%%%%%%%%%%%%%%%%%%%%%%%%
\subsection{Implementation of dynamic contact angle}
\label{dca-implementation}
%%%%%%%%%%%%%%%%%%%---------------------------------%%%%%%%%%%%%%%%%%%%%%%%%%%

\subsubsection{Dynamic contact angle model}
At the three-phase contact line on a solid wall,
the motion of the fluid interface leads to a dynamic change in the contact angle.
Unlike the static contact angle $\theta_s$, 
previous experimental results demonstrate that
the apparent contact angle $\theta_{app}$ depends on the contact line motion, fluid properties, surface properties and surface tension forces \citep{xia2018moving,gao2018drops,shoji2021measurement,shen2024dynamic}.
Its value changes depending on whether the interface advances or recedes: 
when the liquid phase moves towards the gas phase, the advancing contact angle $\theta_a$ appears ($\mathrm{Ca}>0$), whereas the receding contact angle $\theta_r$ occurs at $\mathrm{Ca}<0$.
Therefore, 
it is essential to first determine whether the interface is advancing or receding and then compute the corresponding dynamic contact angle.

To accurately compute the dynamic contact angle, 
we first determine the capillary number $\mathrm{Ca}$ based on the contact line velocity 
$\vec{u}_{cl}$ and use it to classify the interface motion as advancing or receding.
Here the capillary number is re-defined as
\begin{equation}
    %\mathrm{Ca} = \frac{\mu |{u}_{cl}|}{\gamma} \mathrm{sgn}(\vec{u}_{cl}\cdot \vec{e}),
    \mathrm{Ca} = \frac{\mu {u}_{cl}}{\gamma},
    \label{eq-Ca-sgn}
\end{equation}
\noindent
%where the sign of $ \vec{u}_{cl}\cdot \vec{e} $  determines the direction of interface motion.
If $ u_{cl} >0$, 
the interface is advancing with a positive $\mathrm{Ca}$, 
whereas if $ u_{cl} <0$, the interface is receding with a negative $\mathrm{Ca}$.
%The projection of the contact line velocity onto the interface normal in the 
%$xz$-plane is given by
%\begin{equation}
%    \vec{u}_{cl} \cdot \vec{n}_{xz} = % (u_{cl} \cos{\eta})  n_x + (u_{cl} %\sin{\eta}) n_z.
%    \label{eq-ucl-nzx}
%\end{equation}
%\noindent
For a 2D case, the dot product between the contact line velocity and the normal vector 
in eq.~(\ref{eq06})
can be simplified as $u_{cl} = u_{x}$. 

The wetting behavior at real surfaces often exhibits significant dynamic changes of the contact angle, deviating from the static value $\theta_s$,
with the advancing-receding difference ($\theta_a - \theta_r$) defined as contact angle hysteresis (CAH) \citep{gao2006contact,butt2022contact}.
The hysteresis behavior necessitates multiscale modeling: macroscopic viscous bending dominates interface deformation, and microscopic phenomena like molecular adsorption and defect pinning govern contact line mobility.
To unify these mechanisms, 
Dwivedi et al.~\citep{dwivedi2022dynamic} came up with a combined dynamic contact angle model that explicitly incorporates CAH by accounting for both frictional and pinning forces at the microscopic scale.
The formula is expressed as:
%-----------------------------------------
\begin{equation}
    \theta_{app}^3 = \left \{\arccos \left[ \cos{\theta_s}-\frac{\xi_{a/r} \mathrm{Ca}}{\mu } - \frac{C_{pin} \tanh(C \times \mathrm{Ca})}{\gamma} \right]  \right \}^3  + 9\,\mathrm {Ca} \ln{\epsilon}, 
    \label{eq-dca-pin}
\end{equation}
%-----------------------------------------
\noindent
where the contact line friction coefficients $\xi_a$ (advancing) and $\xi_r$ (receding) quantify velocity-dependent energy dissipation at the moving contact line. 
The transition between these two regimes is controlled by the factor $C$ for smoothing CAH
and the pinning coefficient $C_{pin}$, defined as:
\begin{equation}
    C_{pin}= 
    \left\{\begin{matrix}
    \gamma (\cos{\theta_s}-\cos{\theta_a}) \quad \mathrm{Ca}>0 \\
    \gamma (\cos{\theta_r}-\cos{\theta_s})
    \quad \mathrm{Ca}<0
    \end{matrix}\right.
    \label{eq-Cpin}
\end{equation}
\noindent
where the parameter $\epsilon$ characterizes the ratio of macroscopic to microscopic length scale.

In each contact-line cell, 
the dynamic contact angle $\theta_{app}$ is then computed via Eq.~\ref{eq-dca-pin}.
This velocity-dependent model indicates that 
adjacent interfacial cells may have different contact angles based on their local flow conditions.
It is essential to maintain complete synchronization and consistency of the dynamic contact angles across all processors in parallel computations (MPI),
as the accurate reconstruction of the interface depends on it.
%Accurate interface reconstruction depends critically on consistent contact angle data at boundaries.
Mismatched values between two neighbouring processors (e.g., 90$^\circ$ versus 100$^\circ$ for the same boundary cell) may create errors in normal vectors, curvature, and surface tension calculations.
In our implementation, the boundary() function of MPI is used for synchronization to ensure identical boundary cell data across all processors.
%The $boundary()$ function of MPI synchronization preserves global consistency by ensuring identical boundary cell data across processors.
This synchronization also maintains physical fidelity by keeping spatial contact angle variations to originate only from real velocity-dependent calculations rather than parallel communication artifacts.

\subsubsection{The height function method}

\begin{figure}[htbp]
    \centering
    \includegraphics[width=0.55\textwidth]{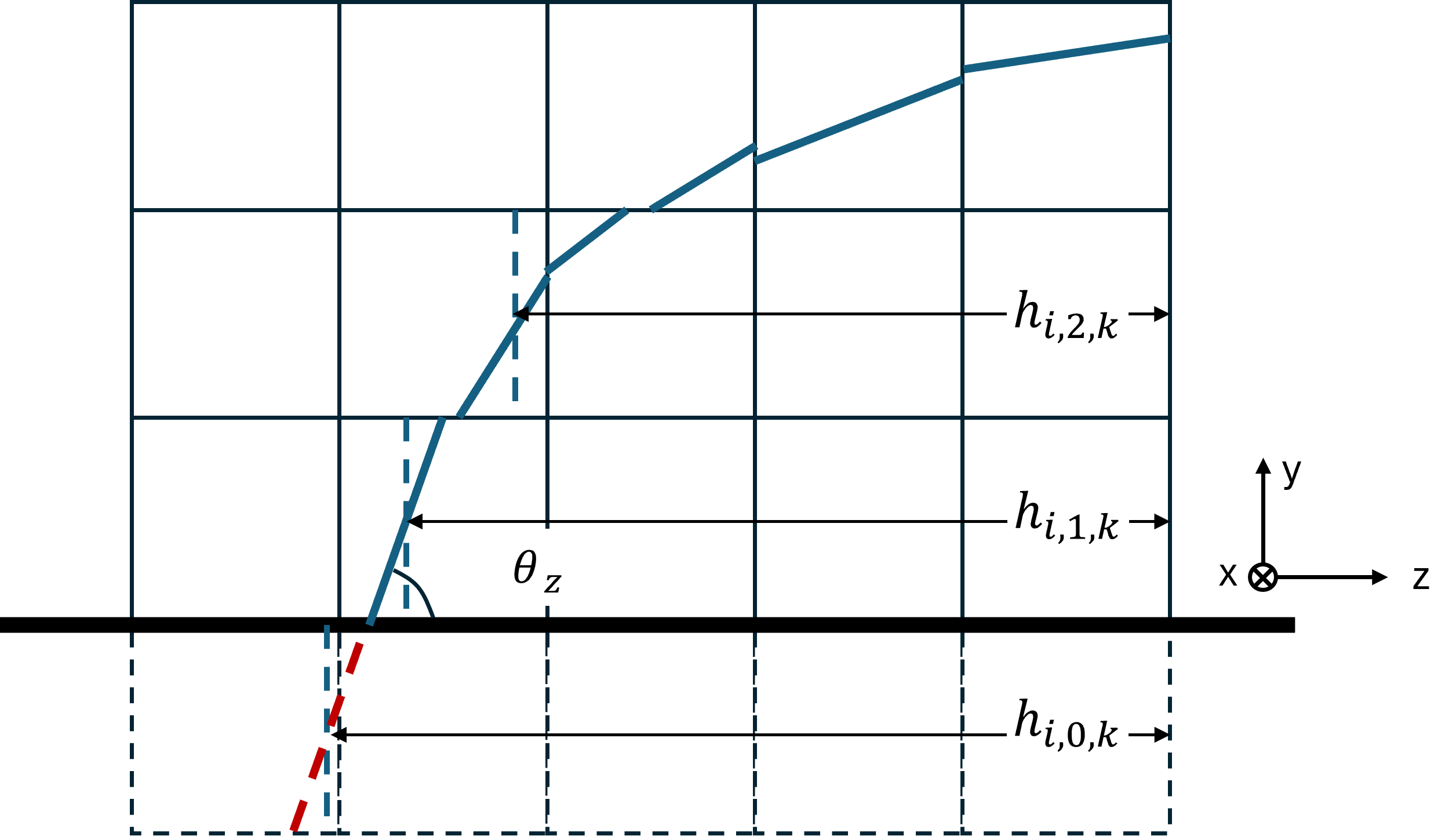}
    \caption{Height functions constructed horizontally in the $yz$-plane. $\theta_{z}$ denotes the projection of the contact angle onto the $z$-axis at the wall.}
    \label{fig-HF3d-yz}
\end{figure}

In Basilisk, 
the contact angle is applied by using the height function (HF) method \citep{afkhami2008height,han2021consistent}. A ghost-cell layer
beneath the surface is used for accurately defining the contact
angle at the surface.
For each interfacial cell ($f\in (0,1)$), the fluid height is computed by summing the volume fractions in the direction most normal to the interface.
For instance, in a 3D case with a primarily horizontal interface, the fluid height in the $y$-direction is written as:
\begin{equation}
    h_{i,k} = \sum_{j}f_{i,j,k}\Delta
    \label{eq-hf}
\end{equation}
\noindent
where $f_{i,j,k}$ denotes the volume fraction of cell ($i,j,k$),
$\Delta$ denotes the uniform grid size,
and $j$ denotes the
cell index in $y$-direction.
Differently, as illustrated in Fig.~\ref{fig-HF3d-yz}, the heights $h_{i,0,k}, h_{i,1,k}, h_{i,2,k}$ are constructed by summing the horizontal fluid columns along the $z$-direction in the $yz$-plane.
Here, the subscript $j=0$ denotes the 
ghost-cell layer below the wall,
where the volume fraction and thus the
height function are calculated
based on the interface 
(red dashed line) that is obtained by linearly extrapolating from the first grid layer above the wall ($j=1$).
%$\theta_{app,z}$ denotes the projection of the apparent contact angle onto the $z$-axis at the wall; the exact relation is provided in the following paragraph.

%--------------------------------------------
%--------------------------

%\textcolor{blue}{Using the ghost-cell layer ensures }

\begin{figure}[htbp]
    \centering
    \includegraphics[width=0.5\textwidth]{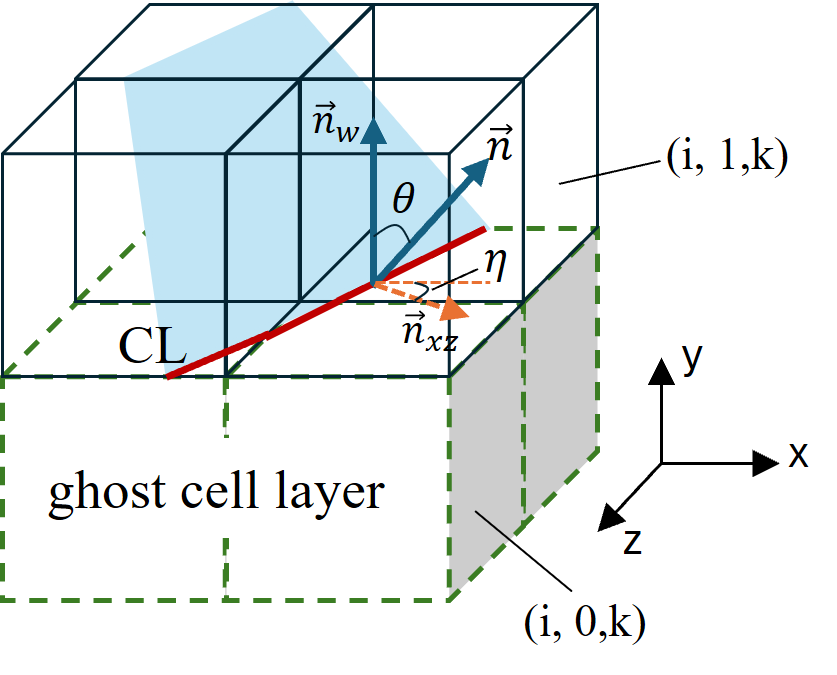}
    \caption{Schematic of a 3D interface intersecting with a solid boundary.}
    \label{fig-HF3d}
\end{figure}

%\st{For a 3D interface intersecting a solid wall, this ghost–cell layer ensures that the reconstructed interface meets the wall at the prescribed apparent contact angle $\theta_{app}$ in each contact–line cell (Fig.~\ref{fig-HF3d}), where $\vec{n}_w$ is the wall-normal unit vector.}
A 3D interface intersecting the solid wall with a contact angle $\theta$ is shown in Fig.~\ref{fig-HF3d},
where $\vec{n}_w$ is the wall-normal unit vector at the interface in the
cell $(i,1,k)$.
%a ghost-cell layer is inserted beneath the wall so that, at every time step, the reconstructed interface meets the wall at the prescribed apparent contact angle $\theta_{app}$ in each contact-line cell (Fig.~\ref{fig-HF3d-yz}), where $\vec{n}_w$ is the unit normal vector to the wall.
Denoting by $\eta$ the angle between the contact–line direction (its projection on the wall) and the chosen HF scan direction, 
the ghost–cell height is prescribed as
\begin{equation}
    h_{i,0,k} = h_{i,1,k} + \Delta / (\tan \theta \cdot \cos \eta)
    \label{eq-ghost-height}
\end{equation}
\noindent
where $h_{i,0,k}$ and $h_{i,1,k}$ represent the fluid height values in the ghost cell and contact-line cell, respectively. 
%--------------------------------------------
%--------------------------
At the current time instant $t^n$, once the updated height functions have been
assembled, the interface geometry on the contact-line cells is evaluated.
In particular, following Afkhami et al.~\citep{afkhami2008height}, 
the curvature in a
contact-line cell is obtained from the ghost-cell height prescribed below the wall.
Using the discrete heights, the (unnormalized) interface normal
$\vec{n}$ and the curvature $\kappa$ are computed as
\begin{align}
    & \vec{n} = (-h_x, 1, -h_z)   \label{eq-normal}\\
    & \kappa = \frac{h_{xx}+h_{zz}+h_{xx}h_z^2 +h_{zz}h_x^2 - 2h_{xz}h_x h_z}{(1+h_x^2+h_z^2)^{3/2}}. \label{eq-kappa} 
\end{align}
\noindent
Here, $h_x, h_z, h_{xx}, h_{zz}, h_{xz}$ denote the first- and second-order partial derivatives of $h$ with respect to $x$ and $z$, approximated using second-order central differences.
For example,
\begin{align}
    & h_{x} = \frac{h_{i+1} - h_{i-1}}{2\Delta} \\
    & h_{xx} = \frac{h_{i+1} - 2 h_i + h_{i-1}}{\Delta^2}
    \label{eq-hx-hxx}
\end{align}
The resulting geometric quantities, $\vec{n}^n$ and
$\kappa^n$, are then supplied to the dynamic-wetting model to re-evaluate the
contact-line velocity for the next time step, $u_{cl}^{\,n+1}$, and to update
the associated boundary values.

%%%%%------------------------------------------------%%%%%%%%%%%%%%%%%%
%%%%%------------------------------------------------%%%%%%%%%%%%%%%%%%
%%%%%------------------------------------------------%%%%%%%%%%%%%%%%%%
%%%%%------------------------------------------------%%%%%%%%%%%%%%%%%%
\section{Droplet impact on solid surfaces}
\label{drop-impact}

%In this section,
%we show three benchmark cases of droplet dynamics on surfaces, each compared with experimental results from the literature  
%to verify the accuracy and efficacy of our numerical methodology.
%The benchmark cases include: droplet spreading on a flat surface, droplet splashing on two substrates, and droplet sliding along inclined planes.
%Each case represents distinct dynamic wetting behaviors,
%enabling comprehensive evaluation of our model's capabilities. 
%Mesh-dependent tests are performed for all cases to ensure numerical reliability.
%The following subsections detail the setup and validation results for each case.

\subsection{Numerical setup}

\begin{figure}
\centering
\includegraphics[width=0.5\textwidth]{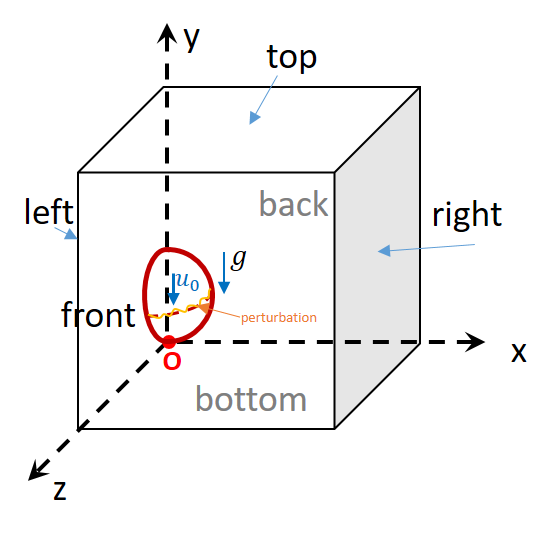}
\caption{\label{fig-3d-sketch-spreading} Schematic of single droplet impact on a solid surface in a 3D computational domain.}
\end{figure}

In this section, we simulate the impact of a single droplet on a solid substrate at
different
impact velocities leading to spreading and splashing.
The computational domain is a cubic box of edge length $5~\mathrm{mm}$ (Cartesian coordinates), 
containing only a quarter of the
volume to reduce computational cost, as shown in
Fig.~\ref{fig-3d-sketch-spreading}. 
A quarter of a spherical droplet 
of a prescribed diameter $R_0$
is initially positioned above a 
horizontal wall 
and moving downwards with an impact velocity $u_0$.
The bottom boundary is the solid wall; the
left and back faces are the symmetry planes, and the top, right, and front faces are treated as pressure/velocity outlets.
A gravitational acceleration of $g$=9.8 m/s$^2$ acts downwards.
The gas-liquid interface in the $xz$-plane is advanced in time, and the apparent contact angle is calculated from
the $xz$-plane during the spreading stage.
For the splashing regime, the setup is identical, except that a small sinusoidal perturbation is superimposed on the initial droplet radius (see Fig.~\ref{fig-3d-sketch-spreading}).

%--------------------------------------------
%-------------------------------------------

The domain is discretized using Basilisk’s octree-based adaptive mesh refinement (AMR) technique to balance accuracy and computational cost. 
The initially uniform grids consist of 32$\times$32$\times$32 cells with the basic refinement level of 5.
In this hierarchical grid system,
the refinement level $L$ follows a binary progression where each subsequent level doubles the spatial resolution of its predecessor.
For example, $L=n$ corresponds to a cell size of $\Delta=W/2^n$,
where $W$ is the length of the cubic computational domain.
The maximum refinement level depends on the specifics of the simulated  problem and is chosen such as to
ensure that the minimum mesh size provides an accurate resolution of the droplet shape and dynamics.
The corresponding resolution
studies are performed in the next section.
The computational time step is adaptively controlled with a maximum value of $10^{-5}$ s, 
selected to balance the numerical accuracy and efficiency.
This value satisfies the the Courant-Friedrichs-Lewy (CFL) stability limit CFL = 0.5.
Each validation case reported in Sections~\ref{drop-impact} and~\ref{drop-slide-NC2023} required between $6{,}000$ and $180{,}000$ CPU-hours of compute time.

All simulations below, including those in Section~\ref{drop-slide-NC2023}, use consistent fluid properties: water ($\rho_l=998~\mathrm{kg\,m^{-3}}$, $\mu_l=1.0\times10^{-3}~\mathrm{Pa\,s}$) as the liquid and air ($\rho_g=1.225~\mathrm{kg\,m^{-3}}$, $\mu_g=1.81\times10^{-5}~\mathrm{Pa\,s}$) as the surrounding gas, with an interfacial tension $\gamma=0.072~\mathrm{N\,m^{-1}}$ at $20^\circ\mathrm{C}$.

%%%%%%%%%%%%%%%%%%%%%%%%%%%%%%%%%%%%%%%%%%%%%%%%%%%%%%%%%%%%%%%
\subsection{Droplet spreading on a treated silicon surface}
\label{drop-spread-POF2009}
\begin{figure}
\centering
\includegraphics[width=0.5\textwidth]{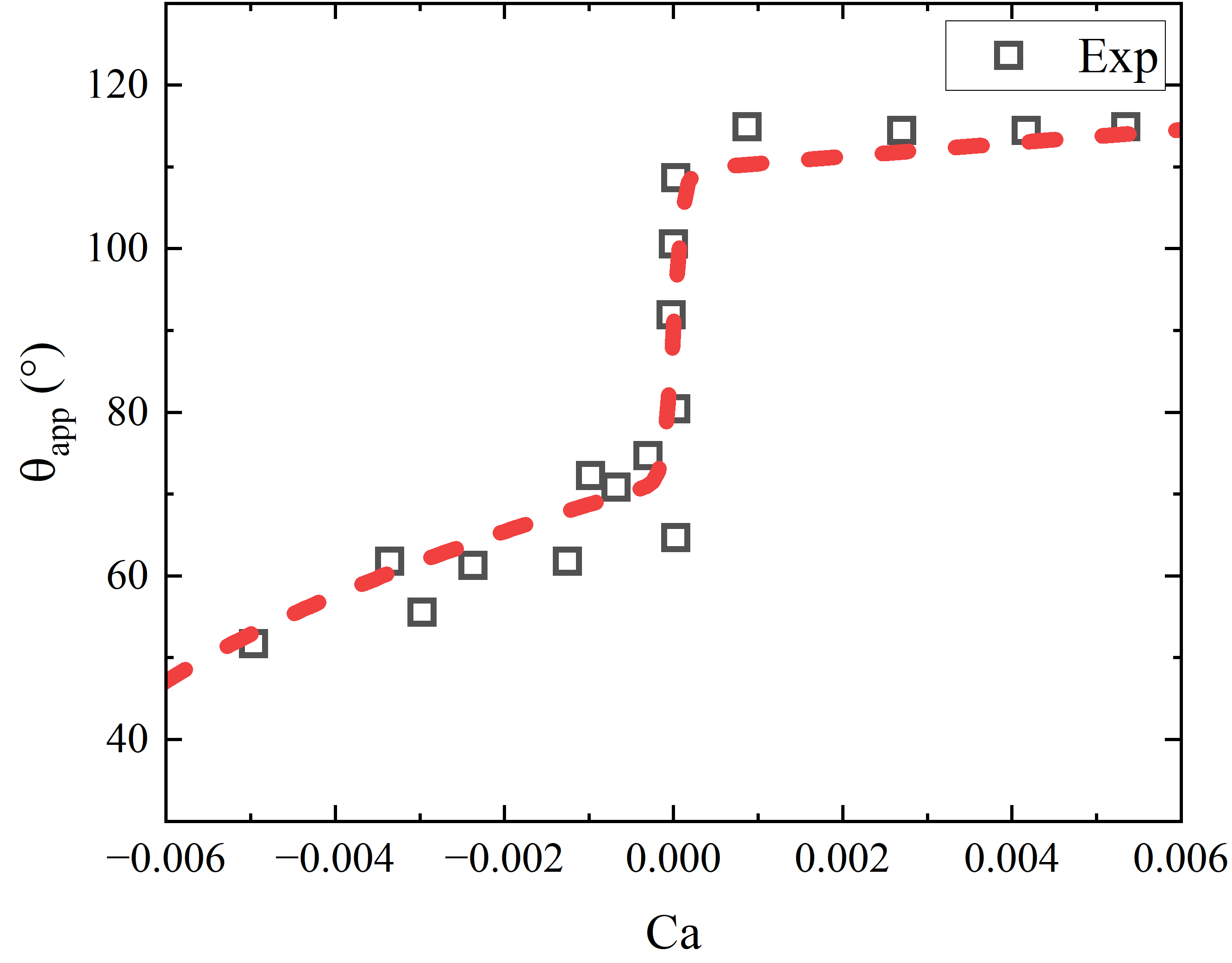}
\caption{\label{fig-dca-theta90} Apparent contact angle versus capillary number on a treated silicon surface: experimental data \citep{yokoi2009numerical} and fitting curve.}
\end{figure}

%%%%%%%%%%%%%%%%%%%%%%%%%%%%%%%%%%%%%%%%%%%%%%%%%%%%%%%%%%%%%%%%%
First, we validate our numerical framework by simulating the three-dimensional spreading dynamics of a single water droplet and comparing with earlier experimental results from Yokoi et al. \citep{yokoi2009numerical}.
%The domain is defined as a cubic box 5 mm in length with six boundaries, as illustrated in Figure \ref{fig-3d-sketch-spreading}.
A quarter droplet with an initial radius of $R_0$=1.14 mm is initially placed above the no-slip bottom boundary with a vertical impact velocity $u_0$=1 m/s.
The corresponding Weber number is We=31.67.
%The left and back sides are set as symmetric boundaries for saving computational effort.
%The top, right and front sides are set as pressure and velocity outlet boundaries.
%A gravitational acceleration $g$=9.8 m/s$^2$ is applied vertically downward along the $y$ direction.
%The gas-liquid interface in the $xz$-plane is tracked dynamically to capture the evolution of the apparent contact angle, as the droplet spreads.
Figure \ref{fig-dca-theta90} presents the
measured wetting dynamics 
of apparent contact angles (denoted by square markers) versus Capillary number
on a treated silicon surface.
A fitting curve (red dashed line) 
is added, that was is computed via Eq.~\ref{eq-dca-pin},
with the corresponding fitting parameters presented in Table.~\ref{tab-case01}.

%--------------------------------------------
%-------------------------------------------

%--------------------------------------------
    \begin{table}[]
    \centering
    \caption{\label{tab-case01}Dynamic wetting parameters of a treated silicon surface.}
    \begin{tabular}{lllllll}
    \hline
              $\theta_s (\mathrm{^\circ})$    &$\theta_a  (\mathrm{^\circ})$  &$\theta_r  (\mathrm{^\circ})$ & $\xi_a  (\mathrm{Pa \cdot s})$ & $\xi_r  (\mathrm{Pa \cdot s})$ & \quad $C$ & \quad$\epsilon$ \\\hline
         90 & 109.5 & 72 & \quad 0.002 & \quad 0.002 & $2\times 10^4$ & $1.0\times 10^7$    \\\hline
    \end{tabular}
    \end{table}
%---------------------------------------------

\begin{figure}
\centering
\includegraphics[width=0.99\textwidth]{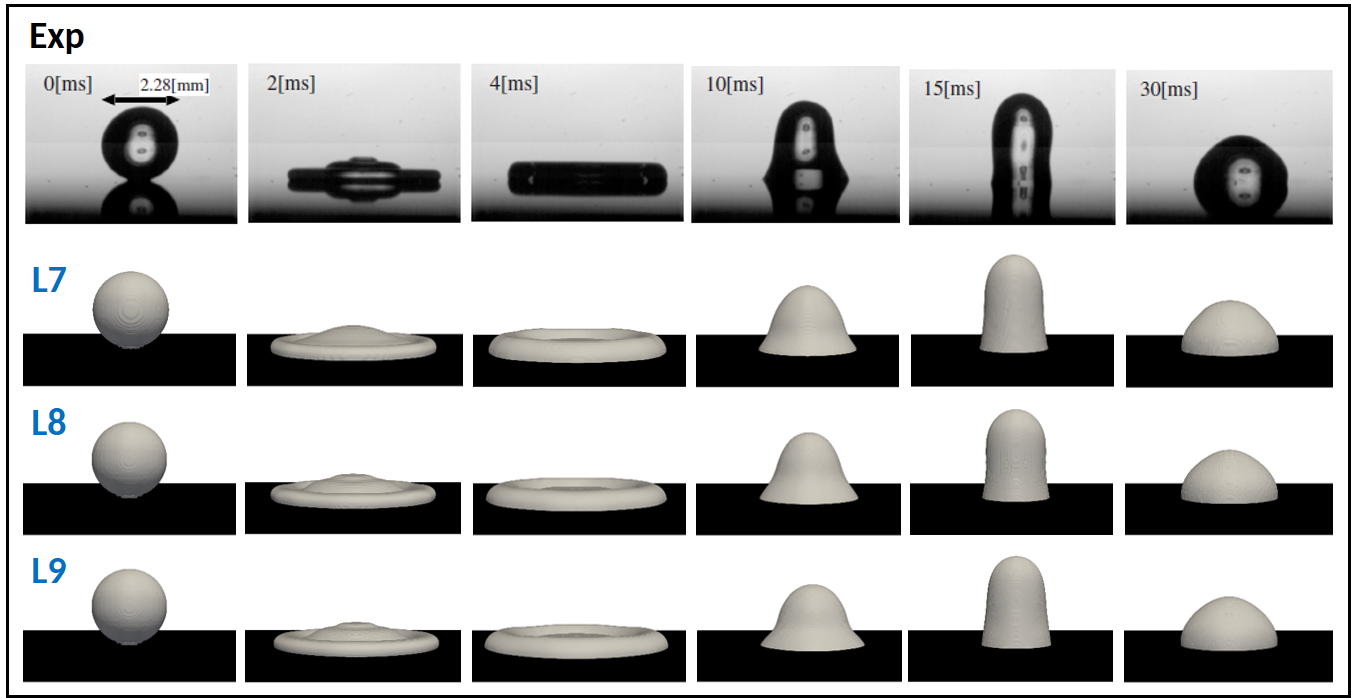}
\caption{\label{fig-3d-photo-Pof2009} Comparison of photographs 
taken during the experiment (top)
and numerically simulated isosurfaces
of a spreading water droplet at selected time instants for different refinement levels L.}
\end{figure}

Figure \ref{fig-3d-photo-Pof2009} presents a visual comparison between snapshots of the experiment
\citep{yokoi2009numerical} and 
corresponding isosurfaces from the
simulations at selected time instants for three mesh refinement levels.
The photographs shown in the top row 
capture the droplet's evolution at key moments—from initial impact (0 ms) through maximum spreading (around 4 ms), subsequent rebound (10–15 ms), and final stabilization (30 ms). 
Below, the corresponding simulation
results are displayed for each refinement level, where L7, L8, and L9 denote minimum mesh sizes of 40 $\mu$m, 20 $\mu$m, and 10 $\mu$m, respectively.
All three numerical cases qualitatively reproduce the overall droplet dynamics.
However, discrepancies are more noticeable at lower refinement levels. 
Specifically, the L7 case shows a slightly more rounded droplet during the spreading stage and a less pronounced vertical elongation during the rebound stage. 
In contrast, both L8 and L9 cases show markedly improved fidelity, thereby
accurately capturing the experimentally observed stepped profile at $t$=2 ms and the characteristic tall, narrow shape during rebound at $t$=15 ms.

%--------------------------------------------
%-------------------------------------------

\begin{figure}
         \centering
         \begin{subfigure}[b]{0.48\textwidth}
             \centering
             \includegraphics[width=\textwidth]{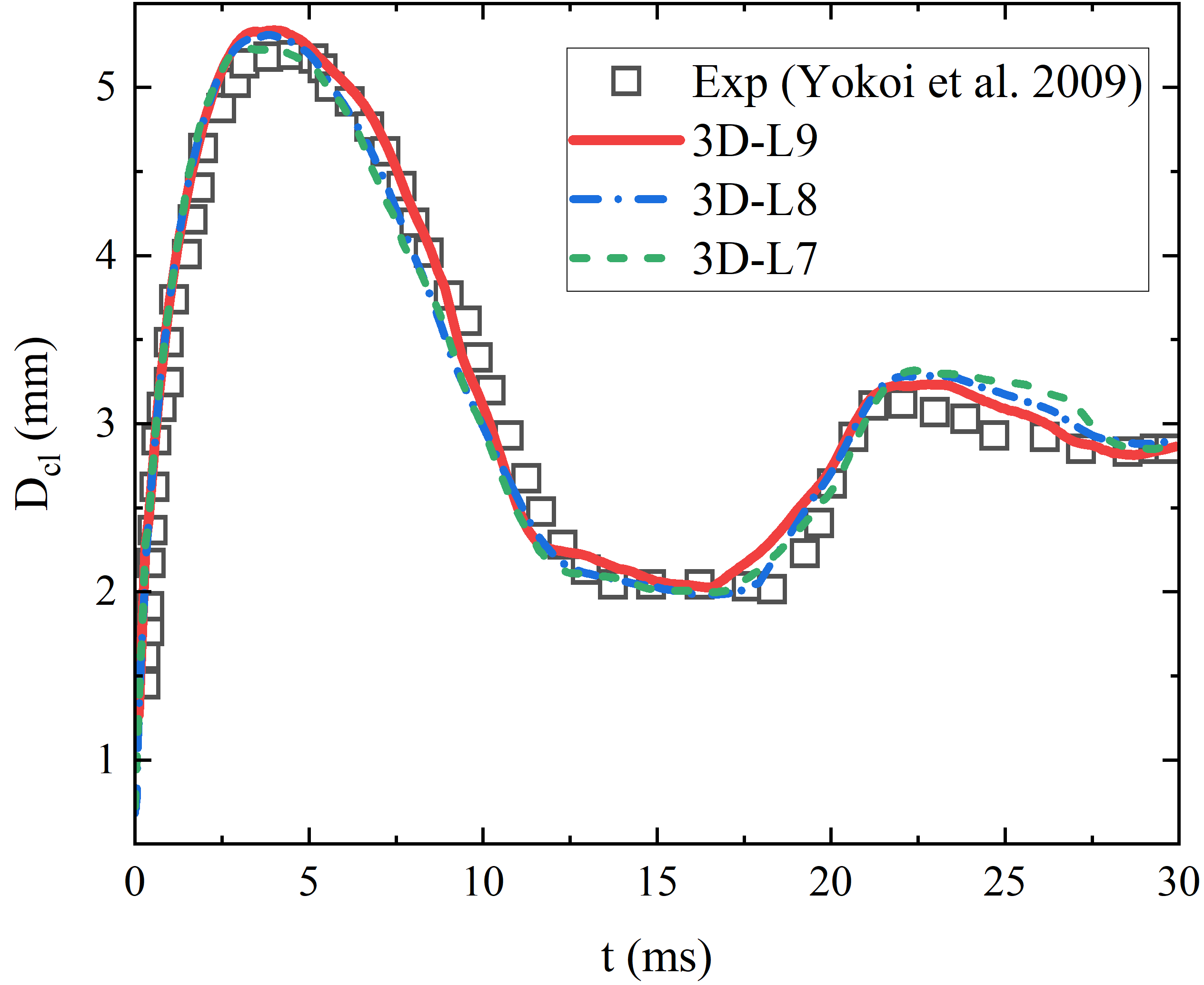}
             \caption{}
             \label{fig-3d-CL-L789}
         \end{subfigure}
         \hfill
         \begin{subfigure}[b]{0.49\textwidth}
             \centering
             \includegraphics[width=\textwidth]{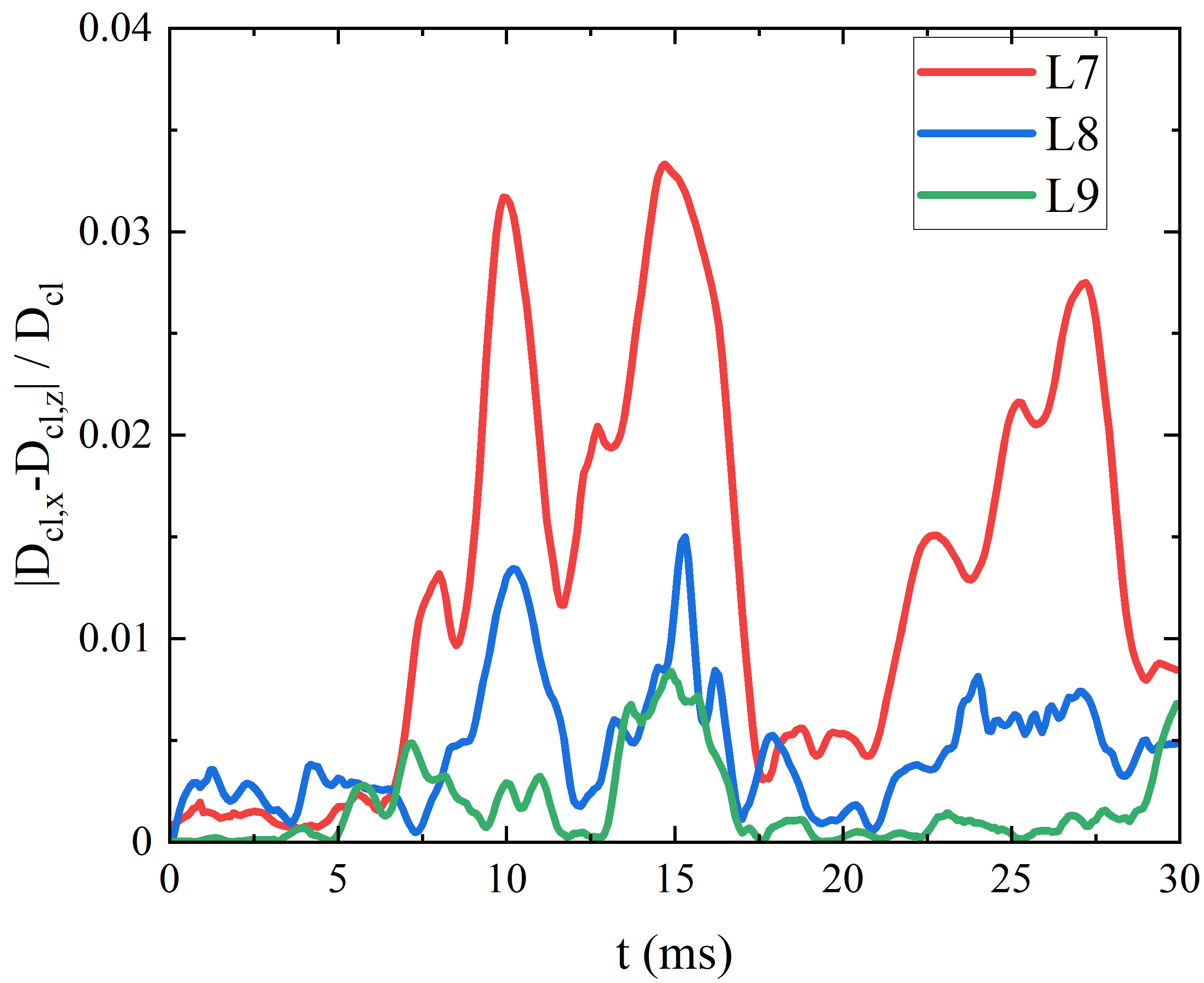}
             \caption{}
             \label{fig-3d-CLerror-L789}
         \end{subfigure}
            \caption{(a) Comparison of the temporal evolution of the contact diameter between experimental \citep{yokoi2009numerical} and simulation results for different refinement levels. (b) Temporal evolution of normalized contact line diameter anisotropy for different refinement levels.}
            \label{fig-3d-pof2009}
    \end{figure}

Figure 
%\ref{fig-3d-CL-L789}
\ref{fig-3d-pof2009}
provides a detailed quantitative comparison of the temporal
evolution of the contact diameter  between experiment and simulations. 
As the experiments report an axisymmetric droplet behavior, here we also check
how accurately axisymmetry is reproduced in our 3D implementation.
Since the droplet is centered at the 
$z$-axis, 
%\st{this definition provides a consistent and geometry-independent measure of the contact line diameter.}
the mean
contact diameter at the wall
is calculated as the average radial distance of the $N_{cl}$ interface points from the droplet center, 
given by
\begin{equation}
    D_{cl} =\frac{2}{N_{cl}} \sum_{i=1}^{N_{cl}}{\sqrt{x_i^2+z_i^2}},
    \label{eq-Dcl}
\end{equation}
\noindent
where $(x_i,z_i)$ are the coordinates of the $i-$th contact line point on the $xz-$plane.
As can be seen in 
Fig.~\ref{fig-3d-CL-L789}, the finest resolution case (L9) demonstrates the best agreement with the experimental results, particularly during the rebound stage.
Importantly, these results are 
consistent also with previous axisymmetric simulations considering
dynamic wetting, as reported in \citep{han2025numerical}.
%\st{, which demonstrated similarly good agreement with experiments under equivalent physical conditions.}
To assess the preservation of axial symmetry, 
Figure \ref{fig-3d-CLerror-L789} shows the temporal evolution of the normalized difference between the contact line diameters in the $x-$ and $z-$directions,
defined as $|D_{cl,x}-D_{cl,z}|/D_{cl}$, for different refinement levels.
%\st{This metric quantifies the degree of deviation from axisymmetry in the 3D simulation. }
Across all cases, the normalized error remains below 4$\%$, and decreases with increasing resolution,
falling below 1$\%$ at L9.
This confirms that the 3D simulation preserves axisymmetry with high fidelity,
thereby validating the robustness of the numerical method in reproducing inherently axisymmetric wetting behavior.

%--------------------------------------------
%--------------------------------------------

\subsection{Droplet splashing on hydrophobic surfaces}
\label{drop-splash-PRL2019}

\begin{figure}
\centering
\includegraphics[width=0.5\textwidth]{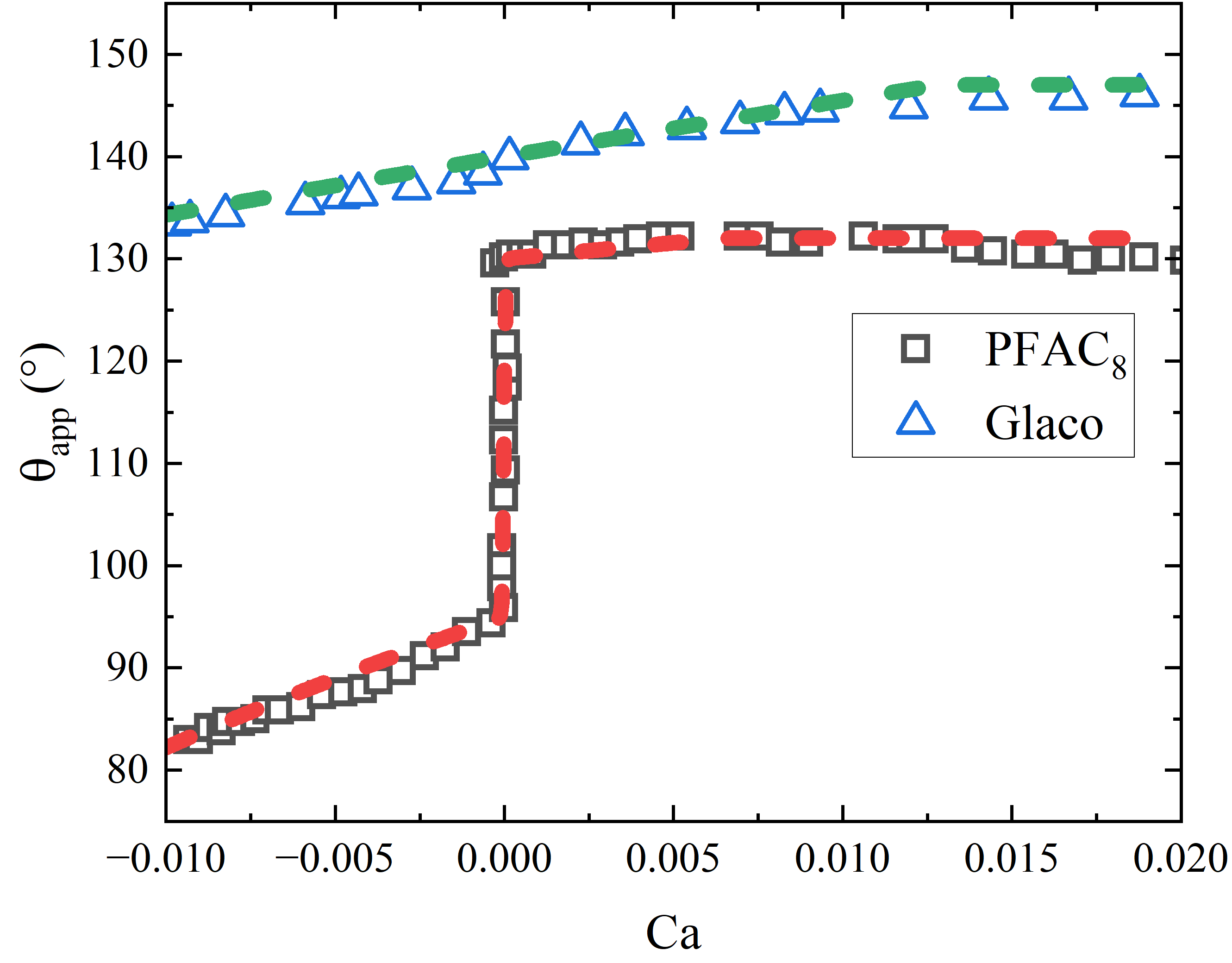}
\caption{\label{fig-dca-PFAC8-Glaco}  Apparent contact angle versus capillary number on two surfaces: experimental data 
            \citep{quetzeri2019role}
            and model prediction. Squares: PFAC$_8$ surface; triangles: Glaco surface.}
\end{figure}

Here, we validate our numerical approach with the experiments of droplet splashing on hydrophobic surfaces \citep{quetzeri2019role}.
The initial radius and impact velocity of the droplet on two surfaces are displayed in Table~\ref{tab-2-Initial-values}.
The corresponding Weber numbers are also listed.
In real experiments, small imperfections or ambient disturbances often lead to the formation of finger-like structures during the splashing process. 
To reproduce this phenomenon numerically, a small sinusoidal perturbation is imposed on the initial droplet radius, as illustrated in Fig.~\ref{fig-3d-sketch-spreading}.
Following Bussmann et al.~\citep{bussmann2000modeling}, we 
initially set
\begin{equation}
    R_{0,p} = R_0 \left[ 1+A_p \cos{\left({2 \pi N}x \right)}\cos{\left({2 \pi N}z \right)} \right],
    \label{eq-perturb}
\end{equation}
\noindent
where $A_p$ represents the amplitude  of the periodic interface perturbation and $N$ denotes its wavenumber.
A small value of $A_p=0.001$ ensures that the perturbation remains in the linear regime.
A choice of $N=4$ introduces a dominant wave number observed in experiments to capture the symmetry-breaking and fingering instability of the interface.
Figure~\ref{fig-dca-PFAC8-Glaco} presents the measured apparent contact angles as a function of the capillary number for two different 
hydrophobic surfaces, PFAC$_8$ and Glaco \citep{quetzeri2019role}.
The dashed curves, calculated using Eq.~\ref{eq-dca-pin},
show good agreement with 
the experimental data when employing the fitting parameters listed in Table~\ref{tab-PFAC8-Glaco}.
Since no contact angle hysteresis was observed on the Glaco surface, 
the factor $C$ is set to unity.

%--------------------------------------------
%-------------------------------------------

%--------------------------------------------
    \begin{table}[]
    \centering
    \caption{\label{tab-2-Initial-values}Initial values of the splashing droplet on two  hydrophobic surfaces.}
    \begin{tabular}{llllllll}
    \hline
                    & $R_0 (\mathrm{mm})$    &$u_0  (\mathrm{m/s})$  &$\mathrm{We}$  \\\hline
        PFAC$_8$    & 1.224                  & 2.34                   & 187   \\\hline
        Glaco       & 1.356                  & 2.09                   & 167   \\\hline
    \end{tabular}
    \end{table}
%---------------------------------------------

%--------------------------------------------
    \begin{table}[]
    \centering
    \caption{\label{tab-PFAC8-Glaco}Dynamic wetting parameters of two  hydrophobic surfaces.}
    \begin{tabular}{llllllll}
    \hline
             & $\theta_s (\mathrm{^\circ})$    &$\theta_a  (\mathrm{^\circ})$  &$\theta_r  (\mathrm{^\circ})$ & $\xi_a  (\mathrm{Pa \cdot s})$ & $\xi_r  (\mathrm{Pa \cdot s})$ & \quad $C$ & \quad$\epsilon$ \\\hline
        PFAC$_8$    & 120& 130 & 95 & \quad 0.0001 & \quad 0.01 & $2\times 10^4$ & $1.0\times 10^4$    \\\hline
        Glaco   & 140 & 147 & 133 & \quad 0.0025 & \quad 0.0025 & 1 & $1.0\times 10^4$    \\\hline
    \end{tabular}
    \end{table}
%---------------------------------------------

\begin{figure}[htbp]
    \centering
    %%%%% 第一行 (a)  %%%%%
    \begin{subfigure}[b]{0.89\textwidth}
        \centering
        \includegraphics[width=\textwidth]{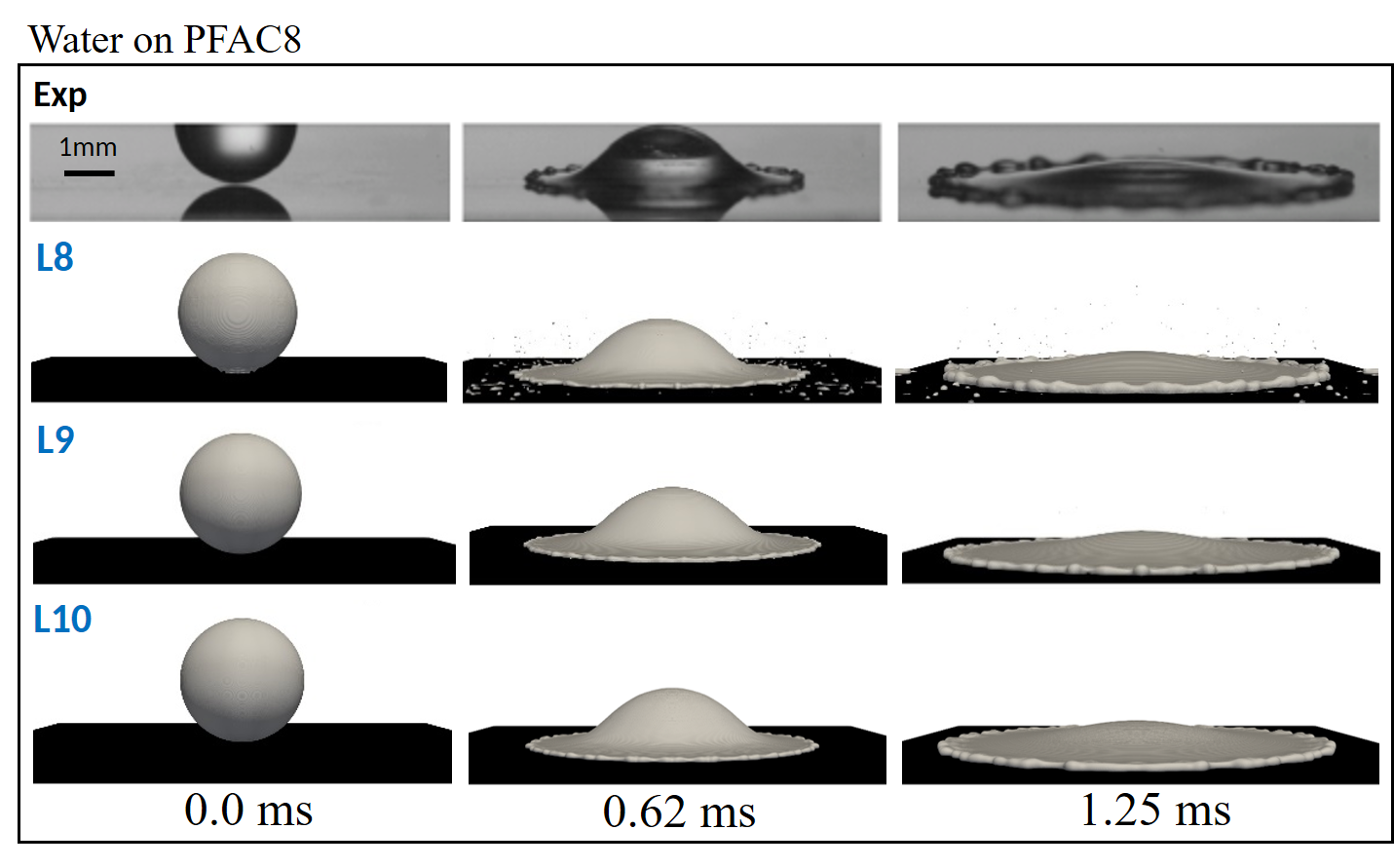}
        \caption{}
        \label{fig-pfac8}
    \end{subfigure}
   
    \vspace{1em} % 可根据需要调整行间距
    
    %%%%% 第二行 (b) %%%%%
    \begin{subfigure}[b]{0.89\textwidth}
        \centering
        \includegraphics[width=\textwidth]{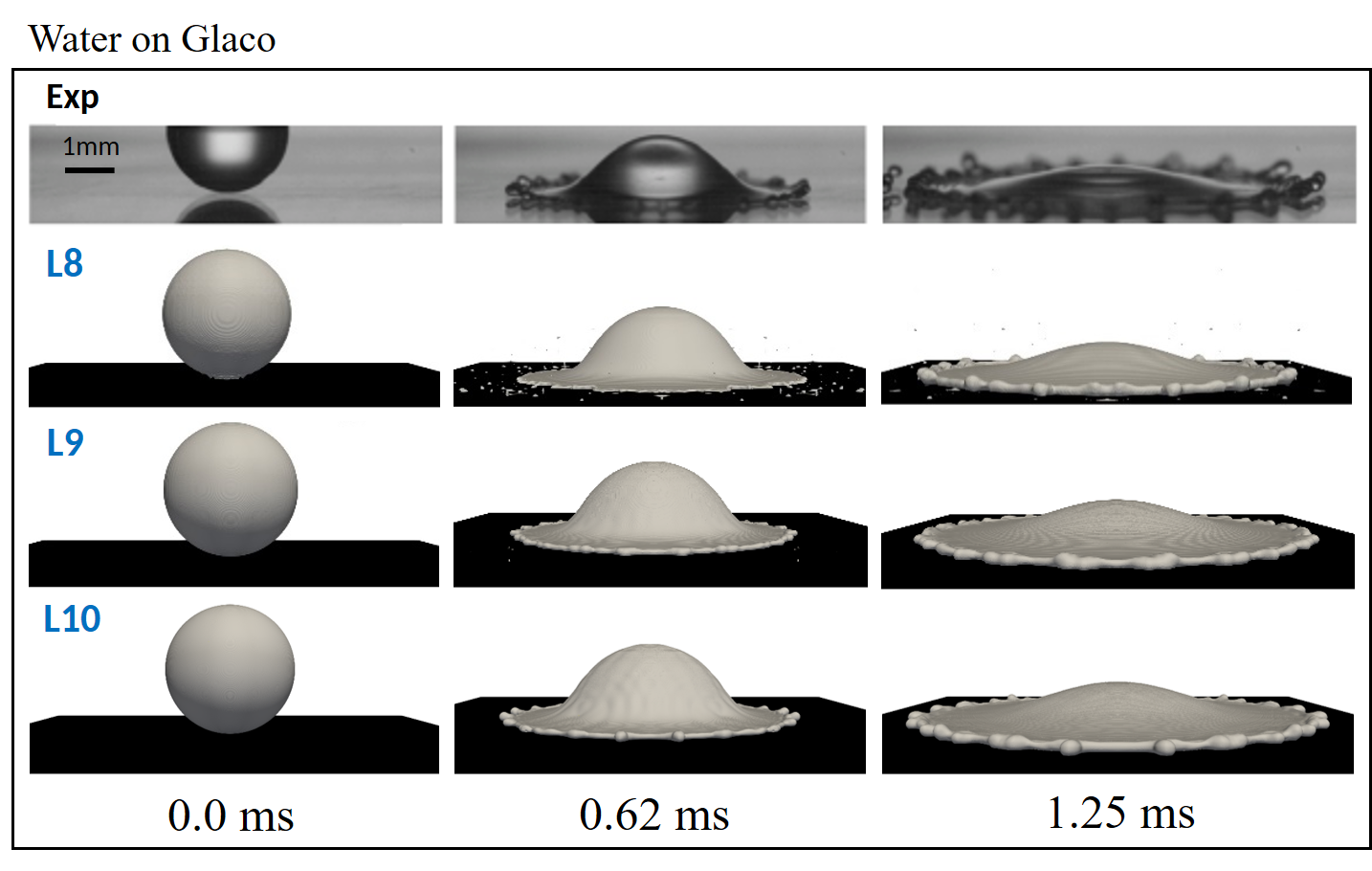}
        \caption{}
        \label{fig-glaco}
    \end{subfigure}
    
    \caption{
        Comparison of photographs (experiment) and isosurfaces
        (simulation)
        for the impact of a
        water droplet on 
                two different substrates: (a) PFAC$_8$, (b) Glaco.
    }
\end{figure}

Figures \ref{fig-pfac8} and \ref{fig-glaco} present comparative analyses between experimental observations and numerical simulations of the impact dynamics of a
water droplet  on the two different
hydrophobic substrates.
Each figure depicts a temporal sequence at 0.0 ms, 0.62 ms, and 1.25 ms, 
capturing the evolution of droplet spreading and splashing. 
The top row in each panel ("Exp") shows snapshots from the experiments, while the subsequent rows (L8, L9, and L10) represent isosurfaces obtained from
numerical simulations
at various refinement levels.
At refinement level
L8, at both surfaces, tiny droplet-shaped numerical artefacts 
are found near the interface, 
which disapper at further grid refinement.
In Figure \ref{fig-pfac8} (PFAC$_8$ surface), 
all three cases (L8-L10) qualitatively 
reproduce the symmetric lamella of the droplet at $t = 0.62$ ms and the development of the fingering structure at $t = 1.25$ ms.
The consistency across the refinement levels suggests that the PFAC$_8$ case is demanding less resolution for capturing the primary features of the impact event.
By contrast, Figure \ref{fig-glaco} illustrates the droplet dynamics on the Glaco surface, where the droplet forms a more pronounced wedge-shaped lamella already at $t = 0.62$ ms, which subsequently promotes the evolution of the fingering pattern seen at $t = 1.25$ ms.
This morphology is characteristic of strongly hydrophobic surfaces.
Therefore, the differences in the lamella shape and the spreading behavior between the both surfaces 
studied here underscore the critical role of surface wettability in dictating droplet dynamics.
%\st{However, at the coarsest refinement level (L8), the numerical simulation fails to fully capture the wedge-shaped structure, indicating a higher sensitivity to mesh resolution on the Glaco surface.}

%--------------------------------------------
%--------------------------------------------

\begin{figure}
\centering
\includegraphics[width=0.5\textwidth]{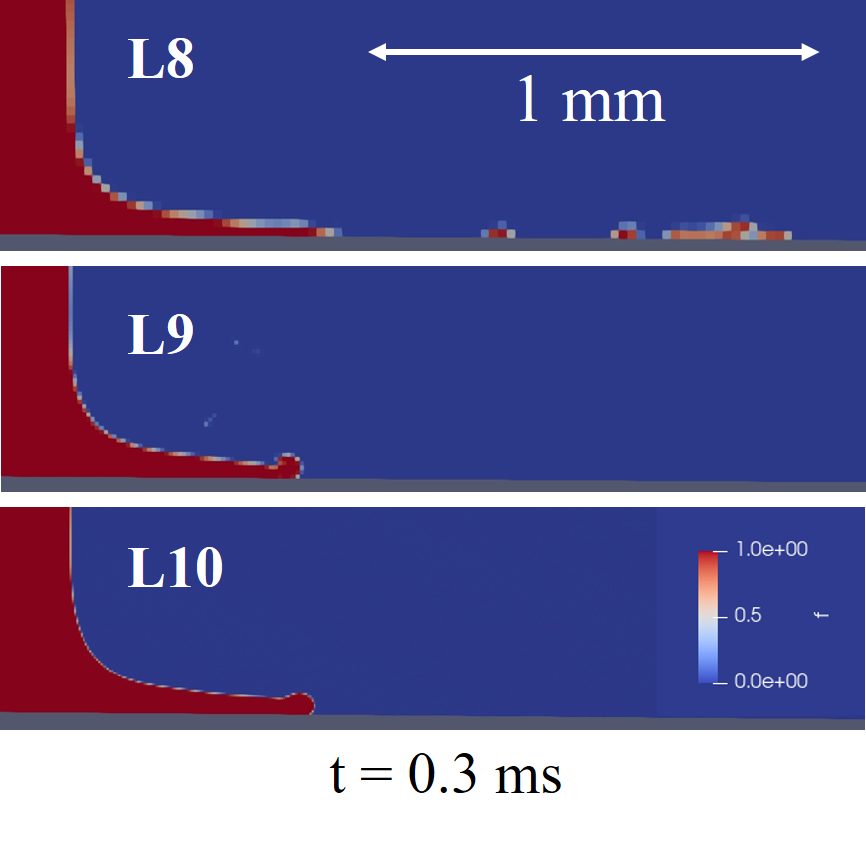}
\caption{\label{fig-t=0.3ms-L8-9-10} Phase fraction of the water droplet on the Glaco surface at different refinement levels at t=0.3 ms along the $x$-axis.}
\end{figure}

Coming back to the interface disturbances observed at refinement
level L8,
Figure~\ref{fig-t=0.3ms-L8-9-10} shows 
the numerically obtained
phase fraction of the water droplet impacting on the Glaco surface at $t = 0.3$ ms for three different levels of mesh refinement: L8, L9, and L10. 
The color scale represents the phase fraction, with red indicating liquid water and blue representing air.
At the lowest refinement level (L8), the interface is rough and shows small broken fragments,
suggesting that the mesh is too coarse to capture the thin liquid layer accurately.
At elevated resolution level L9, the interface becomes smoother, but some small irregularities remain. 
At the highest resolution (L10), the droplet shape is clean and well-defined, with a smooth lamella and a clear interface.
These results highlight the importance of adequate spatial resolution in simulating droplet impact on (super)hydrophobic surfaces, where steep interfacial gradients and thin lamella structures may need to be resolved.

%--------------------------------------------
%----------------------------------------------------------

%----------------------------------------------------------
%----------------------------------------------------------
%----------------------------------------------------------
\section{Droplet sliding on inclined walls}
\label{drop-slide-NC2023}
%----------------------------------------------------------
%----------------------------------------------------------
\subsection{Numerical setup}

\begin{figure}
\centering
\includegraphics[width=0.5\textwidth]{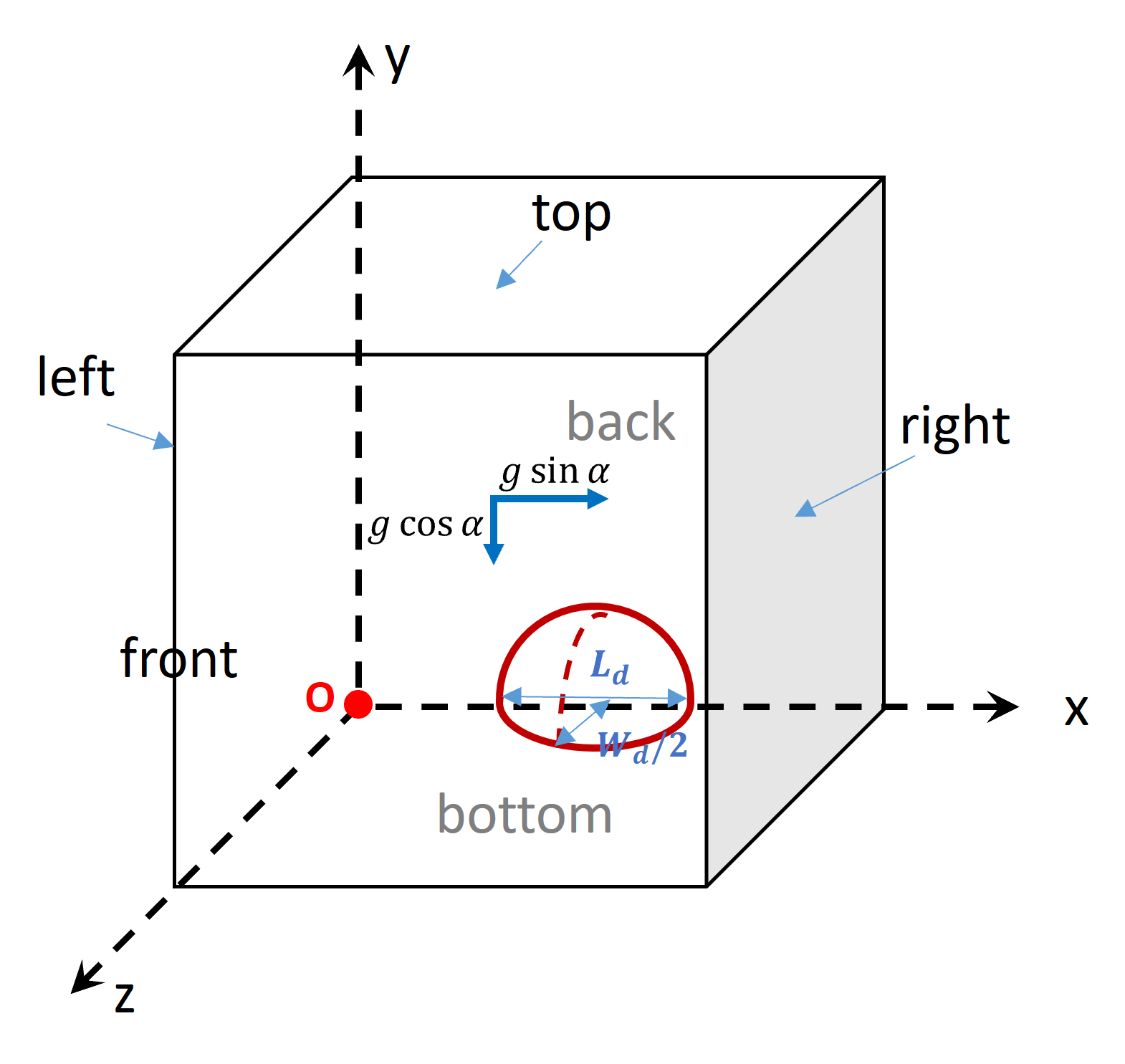}
\caption{\label{fig-3d-sketch-sliding} 3D Schematic of the droplet sliding on an inclined surface in the computational domain.}
\end{figure}

Here
we model the gravity-driven sliding motion of a water droplet on an inclined surface and compare the simulations with experiments reported in Ref.~\citep{li2023kinetic}.
Our simulations aim to accurately capture the sliding dynamics and 
interface deformation by accounting
for dynamic wetting effects.
Figure \ref{fig-3d-sketch-sliding} shows a 0.05 m $\times$ 0.05 m $\times$ 0.05 m computational domain, 
where half of 
a hemispherical droplet, initially at rest, is placed on the lower  wall.
The back face is treated as a symmetry plane to reduce computational cost, while the remaining lateral and top faces are outflow (pressure/velocity outlet) boundaries. 
The bottom wall with a no-slip condition represents the 
sliding solid surface.
To enable sliding at the horizontal
wall, the vector of gravity $\vec{g}$
is inclined by an angle $\alpha$.
Then, the tangential component $g\sin{\alpha}$ along the $x$-axis is driving the downslope motion.
Unfortunately, the referenced experimental report does not specify the exact initial droplet sizes.
Based on calculations from droplet image data on a PS-gold surface and on numerical data used in this reference,
we initialize all
our simulations with a spherical droplet with a radius of $R_0=2.5~\mathrm{mm}$ and the 
corresponding
static contact angle $\theta_s$ 
of the surface (Fig.~13). 
This is different from the initial
situation in the experiment, where
the droplets placed on the inclined surface may deform and accelerate before they are completely released by withdrawing the feeding
needle.
The instantaneous contact length and width
during the sliding will be denoted by $L_d$ and $W_d$, respectively.

%--------------------------------------------
%--------------------------------------------

\begin{figure}
         \centering
         \begin{subfigure}[b]{0.32\textwidth}
             \centering
             \includegraphics[width=\textwidth]{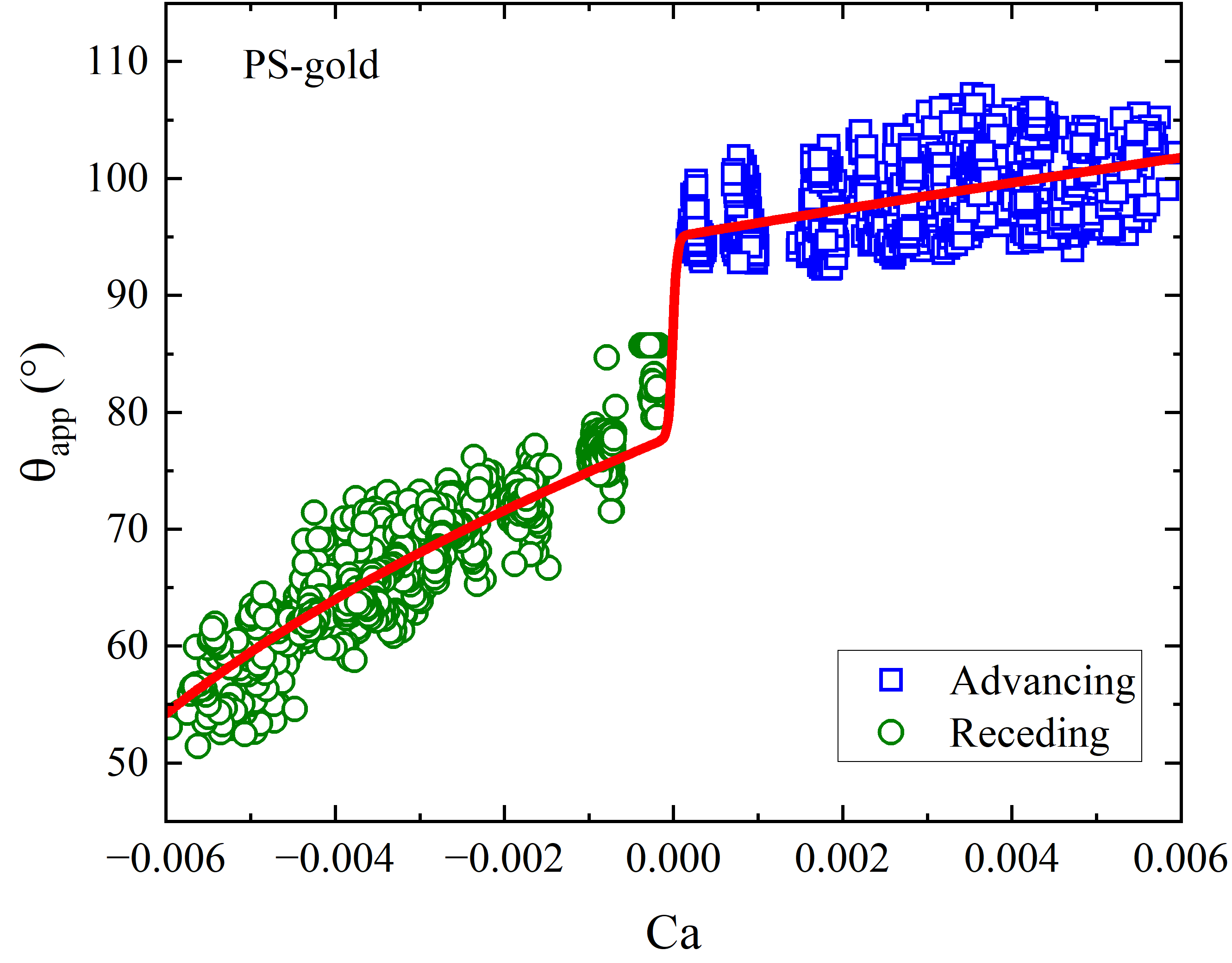}
             \caption{}
             \label{fig-ps-gold}
         \end{subfigure}
         \hfill
         \begin{subfigure}[b]{0.32\textwidth}
             \centering
             \includegraphics[width=\textwidth]{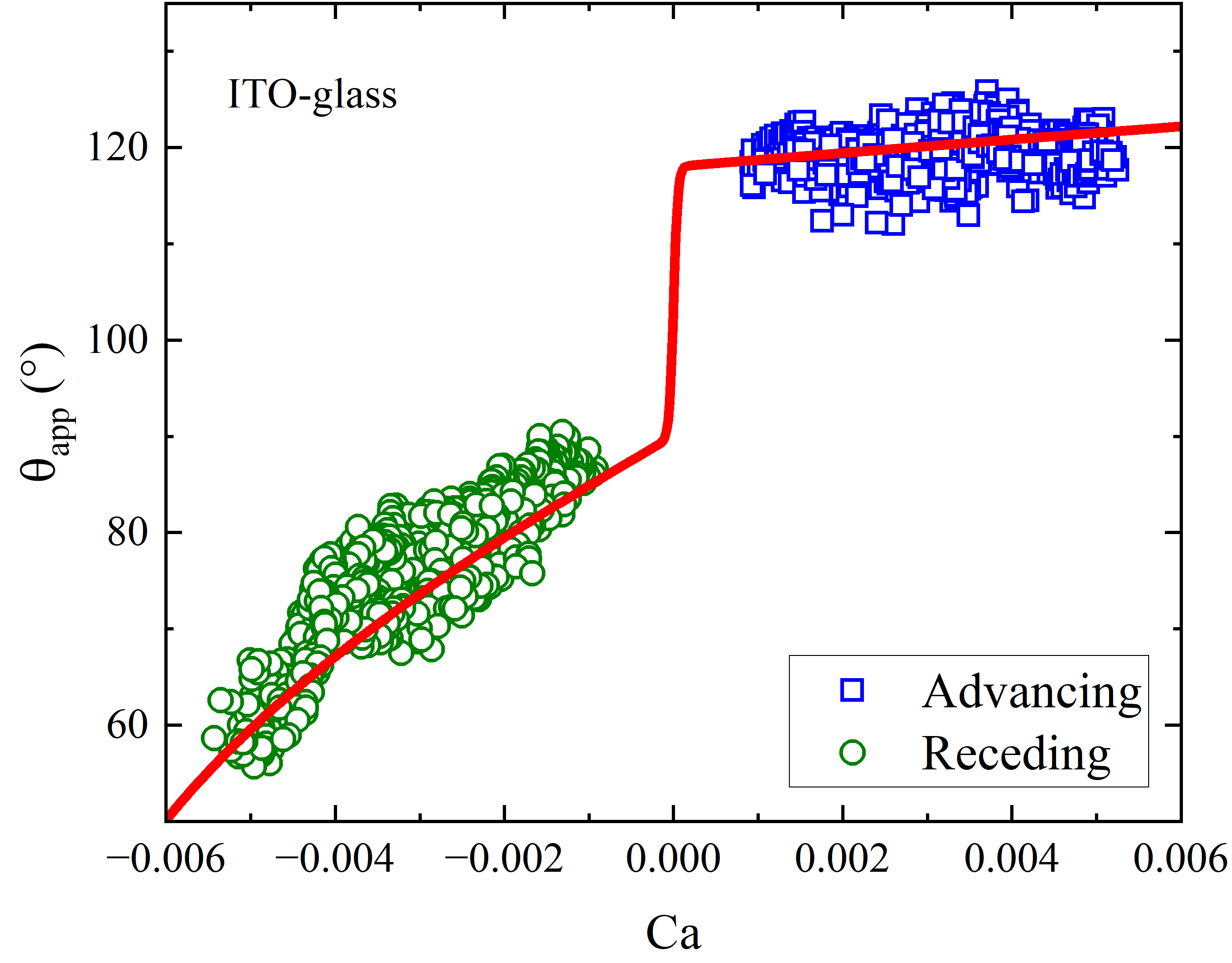}
             \caption{}
             \label{fig-ito-glass}
         \end{subfigure}
                  \hfill
         \begin{subfigure}[b]{0.31\textwidth}
             \centering
             \includegraphics[width=\textwidth]{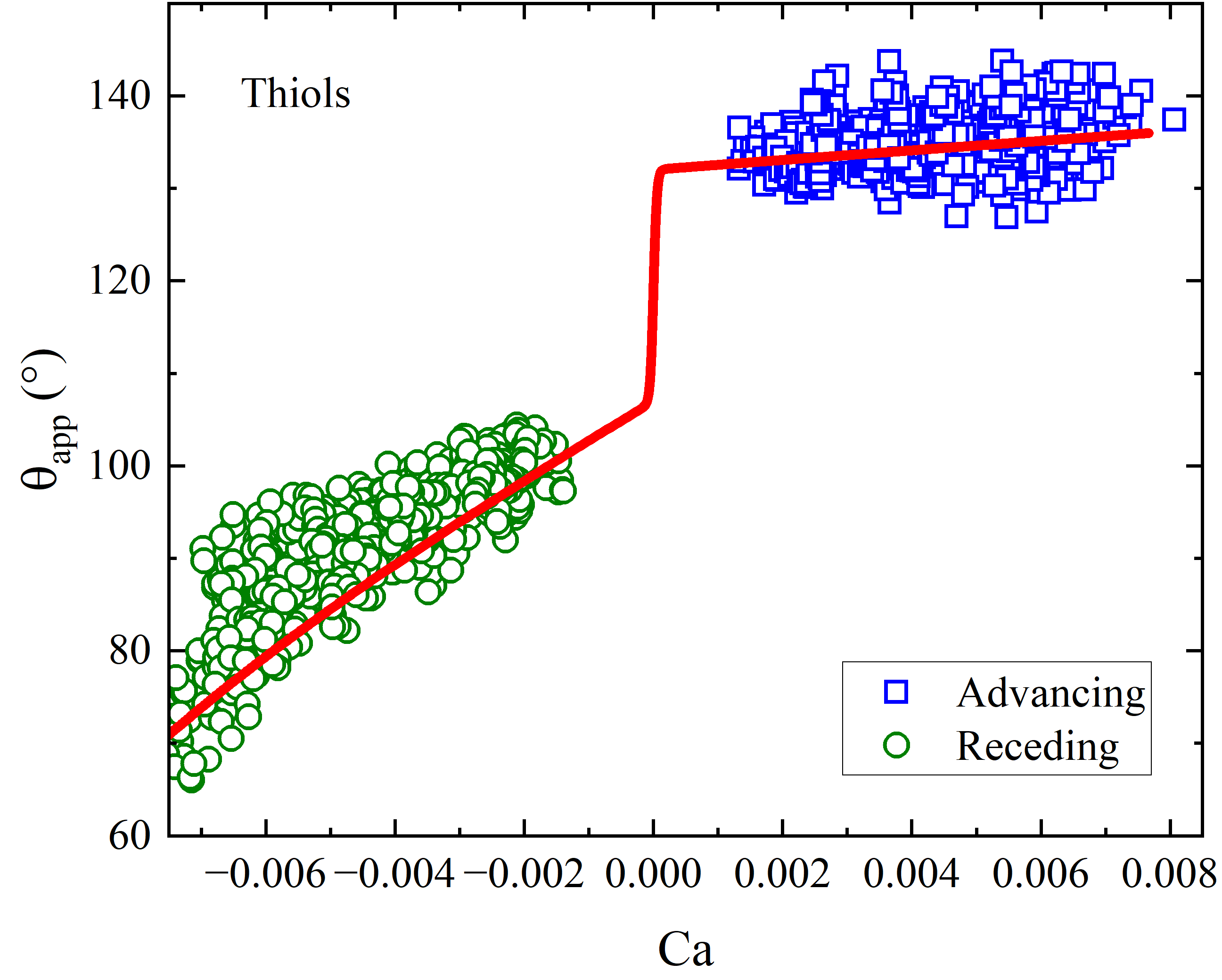}
             \caption{}
             \label{fig-Thiols}
         \end{subfigure}
            \caption{Apparent contact angle versus capillary number on three surfaces: experimental data and model prediction. (a) PS-gold surface; (b) ITO-glass surface; (c) Thiols surface.}
            \label{fig-dca-NC}
    \end{figure}

To validate our numerical approach, we select three characteristic substrates from Ref.~\citep{li2023kinetic}, 
spanning a range of wetting behaviors: 
PS-gold ($\theta_s=88^\circ$), ITO-glass ($\theta_s=104^\circ$), and Thiols ($\theta_s=120^\circ$). 
Figure \ref{fig-dca-NC} shows
measurement data of the dependence of the apparent dynamic contact angle on the capillary number for these surfaces~\citep{li2023kinetic}.
As can be seen, the experimental data show some scatter, which is likely to be caused by surface inhomogeneities.
Based on these data, 
by using Eq.~(\ref{eq-dca-pin}), fit functions for each surface can be obtained,
and the fitted dynamic wetting parameters are
given in Table \ref{tab-NC-dca}.
%\st{analytically solve Eq.~(ref-eq-dca-pin) to reproduce the experimental trends shown as solid curves.}
These fit functions
are superimposed as red lines in Figure \ref{fig-dca-NC}. As can be seen,
the wetting dynamics 
of all three surfaces
is well reproduced.

%This model is implemented at the bottom boundary to govern the dynamic contact angle behavior.

%--------------------------------------------
%--------------------------------------------

%--------------------------------------------
    \begin{table}[]
    \centering
    \caption{\label{tab-NC-dca}Dynamic wetting parameters of three surfaces.}
    \begin{tabular}{llllllll}
    \hline
             & $\theta_s (\mathrm{^\circ})$    &$\theta_a  (\mathrm{^\circ})$  &$\theta_r  (\mathrm{^\circ})$ & $\xi_a  (\mathrm{Pa \cdot s})$ & $\xi_r  (\mathrm{Pa \cdot s})$ & \quad $C$ & \quad$\epsilon$ \\\hline
        PS-gold    & 88 & 95 & 78 & \quad 0.001 & \quad 0.02 & $2\times 10^4$ & $1.0\times 10^8$    \\\hline
        ITO-glass   & 104 & 118 & 90 & \quad 0.001 & \quad 0.062 & $2\times 10^4$ & $1.0\times 10^7$    \\\hline
        Thiols   & 120 & 132 & 107 & \quad 0.001 & \quad 0.055 & $2\times 10^4$ & $1.0\times 10^6$    \\\hline
    \end{tabular}
    \end{table}
%---------------------------------------------

\subsection{Mesh independence}
\label{mesh-independence}

\begin{figure}
         \centering
         \begin{subfigure}[b]{0.48\textwidth}
             \centering
             \includegraphics[width=\textwidth]{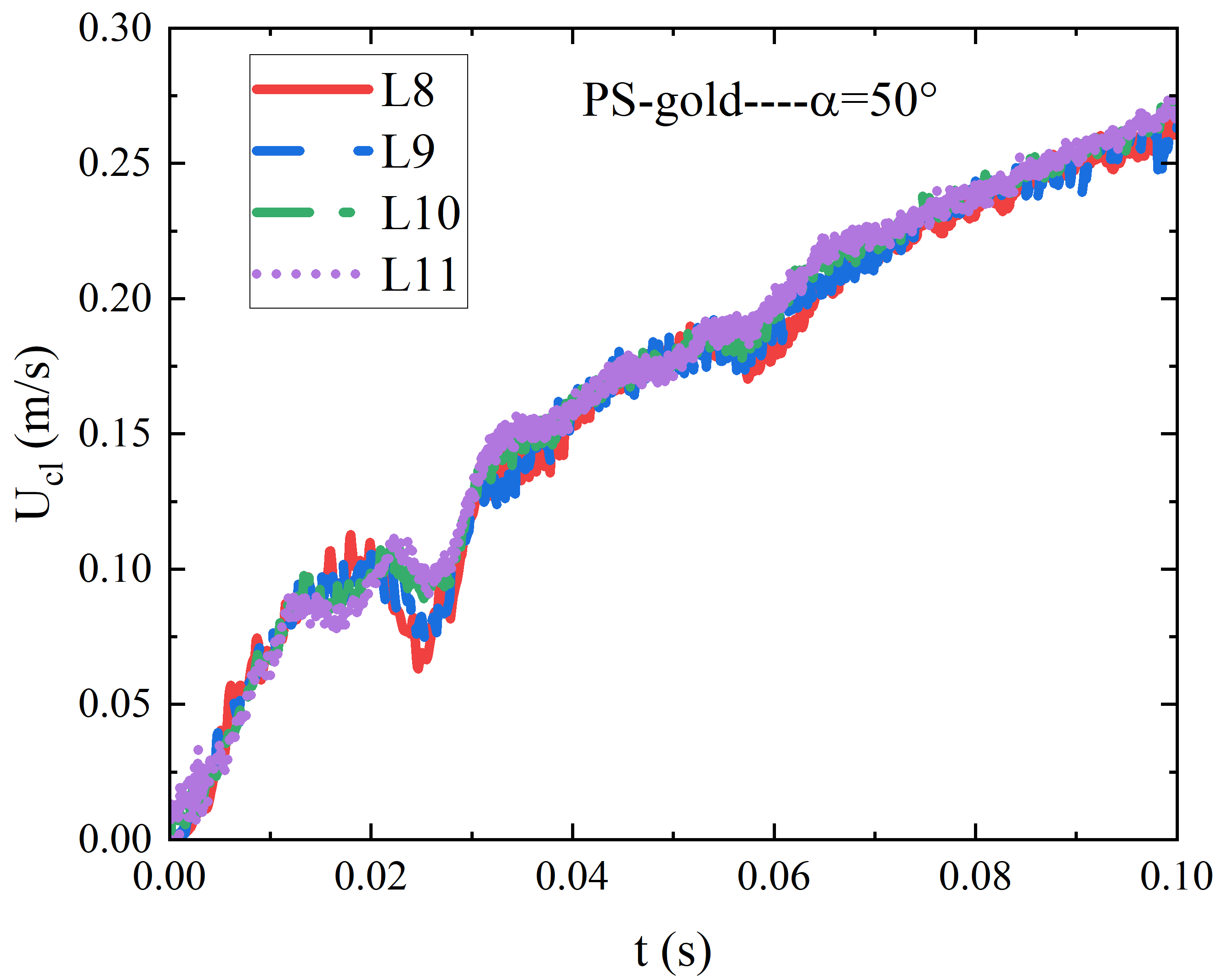}
             \caption{}
             \label{fig-mesh-ucl}
         \end{subfigure}
         \hfill
         \begin{subfigure}[b]{0.46\textwidth}
             \centering
             \includegraphics[width=\textwidth]{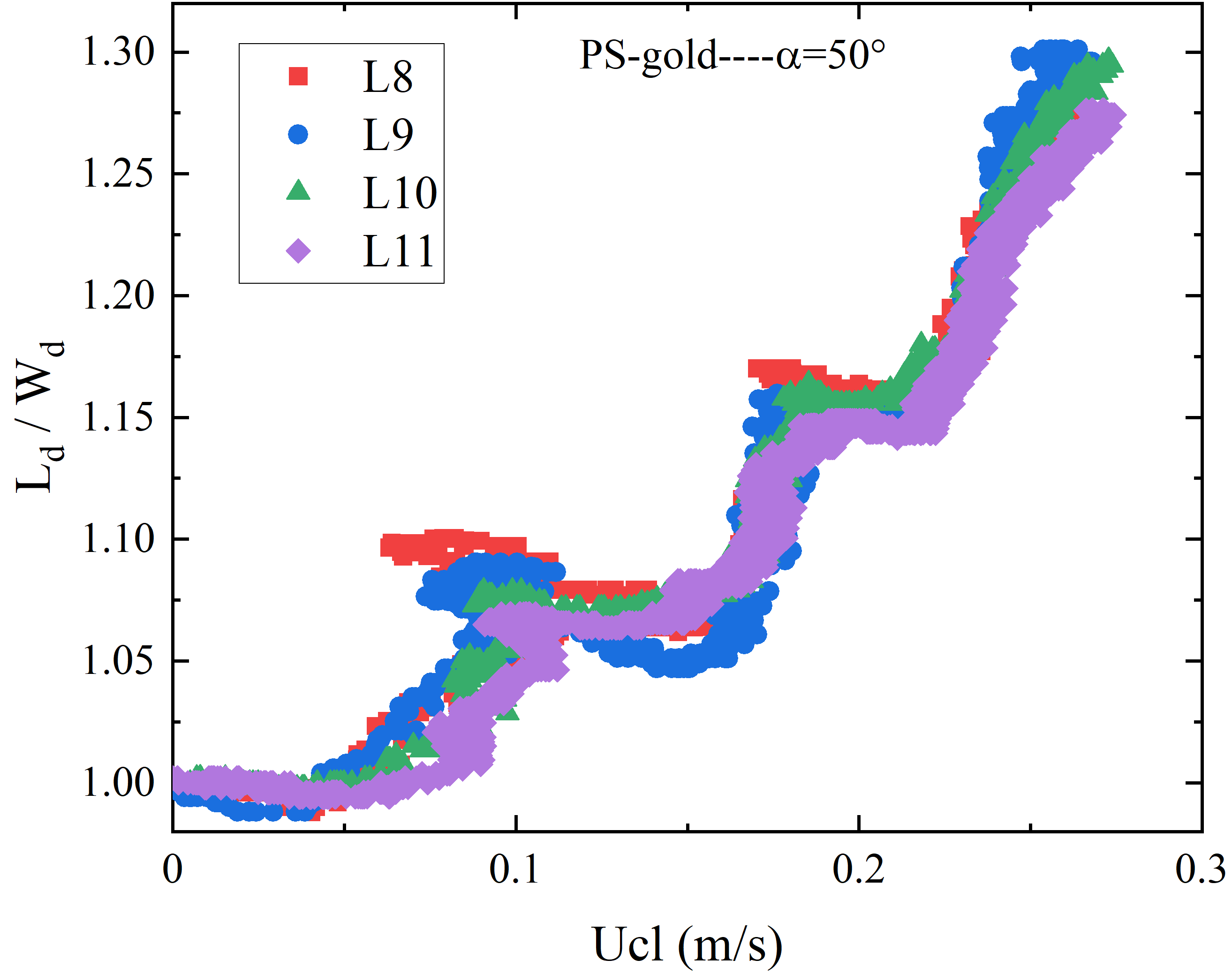}
             \caption{}
             \label{fig-mesh-ratio}
         \end{subfigure}
            \caption{Mesh-resolution analysis: temporal evolution of
            (a) the contact line velocity and (b) the 
            droplet aspect ratio (length-to-width: $L_d/W_d$) on a PS-gold surface at
            $\alpha$=50$^\circ$ 
            .}
            \label{fig-mehsh-NC}
    \end{figure}

Figure \ref{fig-mehsh-NC} presents a 
mesh refinement study
%mesh independent study 
to assess the numerical sensitivity of the simulations for a
sliding droplet on a PS-gold surface inclined by $\alpha=50^\circ$. 
Two key parameters are compared across four refinement levels (L8–L11): (a) the evolution of the average contact line velocity $U_{cl}$ over time, and (b) the droplet aspect ratio $L_d/W_d$, which sensitively depends on the sliding velocity.
Here, $U_{cl}$ is defined as the mean value of the front and back velocities
of the center contact points 
($z=0$) of the droplet.
As can be seen in Figure \ref{fig-mesh-ucl}, all refinement levels (L8–L11) capture the increasing trend of $U_{cl}$ over time, with close agreement among the curves. 
The velocity increases in a stepwise manner over time, with alternating long phases of acceleration and brief phases of deceleration or plateaus of nearly constant velocity. 
This behavior reflects the stick-slip-like dynamics of the contact line, driven by the balance between gravitational forcing and contact angle hysteresis.
Small deviations are observed at the early transient stage (e.g., $t<$ 0.03 s), with coarser meshes (L8, L9) exhibiting more pronounced discrepancies. 
Finer meshes (L10, L11) are found
to better resolve the transient accelerations. 
Despite these differences, 
the convergence across all levels beyond 
$t\approx$ 0.03 s suggests that the averaged contact line velocity becomes mesh-independent at later times
for all refinement levels tested.

Figure \ref{fig-mesh-ratio} presents the correlation between the droplet aspect ratio  $L_d/W_d$ and the contact line velocity $U_{cl}$.
As $U_{cl}$ increases,
the droplet elongates progressively in the flow direction, demonstrating a strong coupling between contact line motion and droplet deformation. The relationship exhibits a staircase-like structure, indicating that elongation occurs in discrete stages corresponding to changes in $U_{cl}$, which again highlights the influence of contact angle hysteresis and interface dynamics.
All refinement levels qualitatively
reproduce the same deformation trend,
and the close overlap between the results for L10 and L11 further confirms numerical convergence.
Together, the figures reveal that the sliding droplet undergoes nonlinear acceleration governed by both gravitational and capillary forces. The interplay between contact line mobility and droplet deformation governs the rate of acceleration, and the consistent trends across refinement levels confirm the robustness of the numerical model.
%Together, the results demonstrate consistent simulation outcomes across refinement levels, with higher mesh refinement level providing enhanced accuracy in resolving droplet dynamics.
In the following subsections, we use the refinement level L10 
%\st{to validate the simulation results.}
in the numerical simulations
for comparison with experimental
results.

%--------------------------------------------
%--------------------------------------------

\subsection{Droplet motion on different surfaces}
\label{exp-3cases-compare}

\begin{figure}[htbp]
    \centering
    % ===== 左侧 minipage，包含 (a) 和 (b) 两个子图 =====
    \begin{minipage}[c]{0.38\textwidth}
        % -- 子图 (a) --
        \begin{subfigure}[c]{\textwidth}
            \centering
            \includegraphics[width=\textwidth]{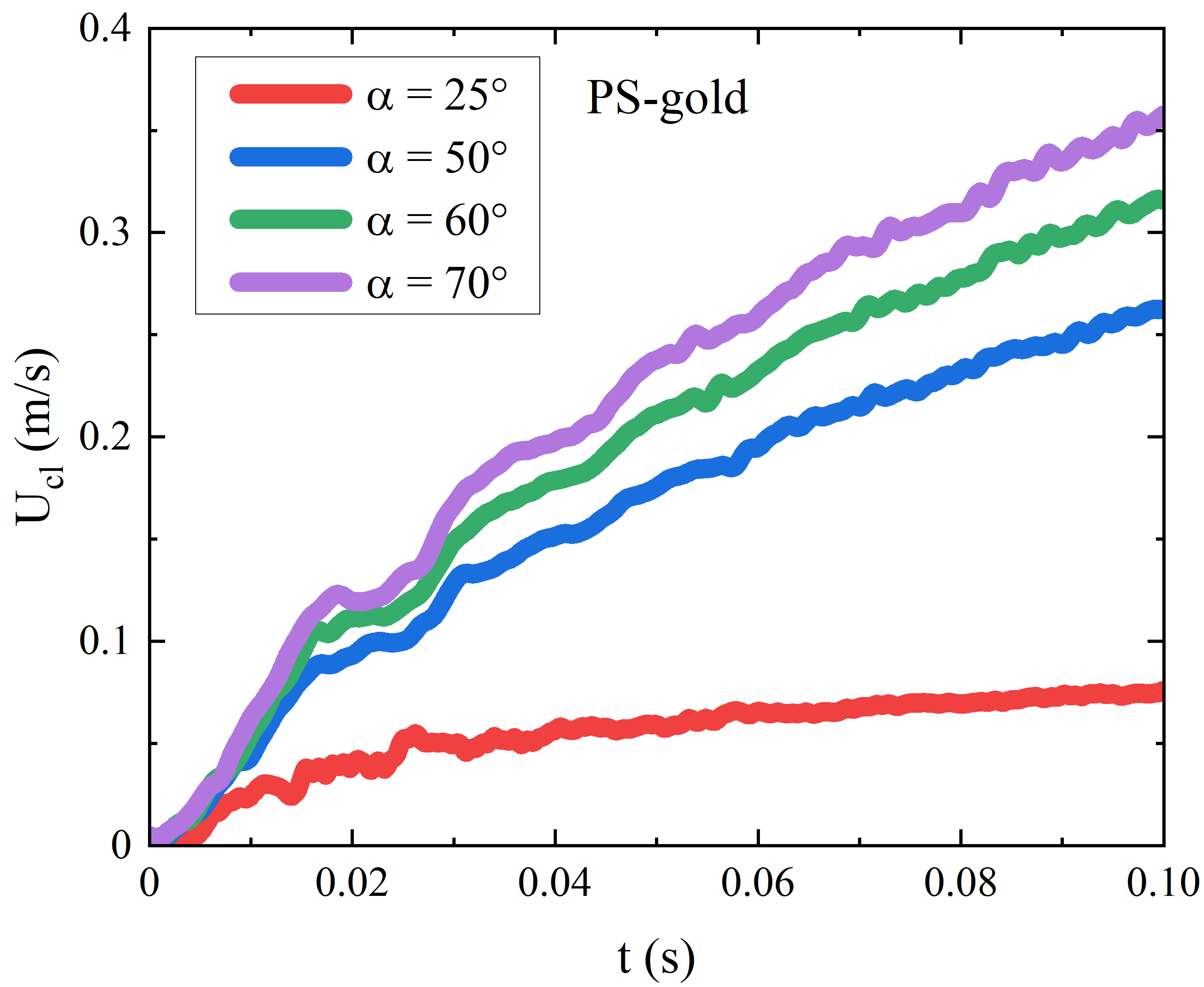}
            \caption{}
            \label{fig-PS-ucl}
        \end{subfigure}
        \vspace{0.5cm}
        % -- 子图 (b) --
        \begin{subfigure}[c]{\textwidth}
            \centering
            \includegraphics[width=0.96\textwidth]{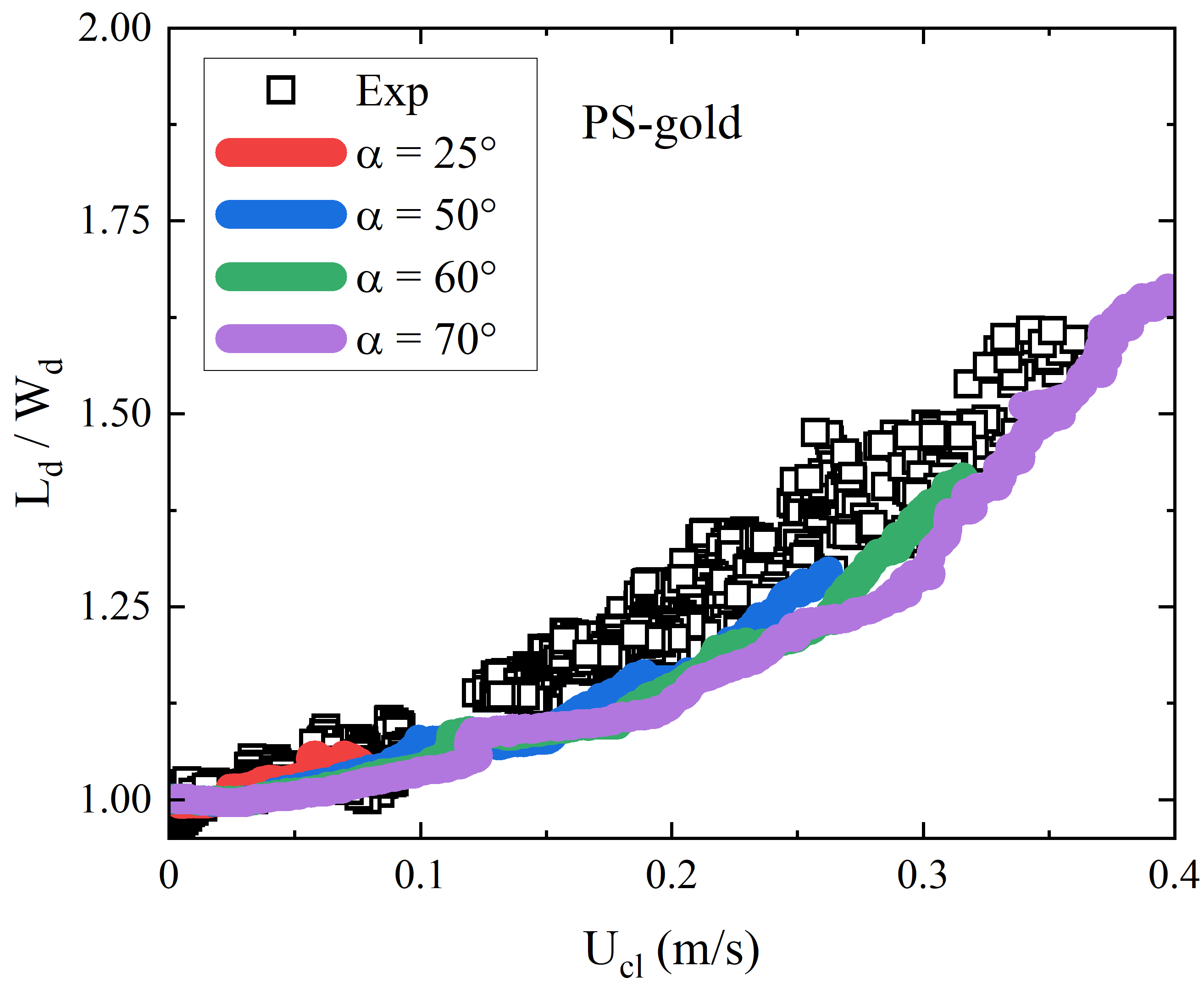}
            \caption{}
            \label{fig-PS-LW}
        \end{subfigure}
    \end{minipage}
    \hfill
    % ===== 右侧子图 (c) =====
    \begin{subfigure}[c]{0.6\textwidth}
        \centering
        \includegraphics[width=\textwidth]{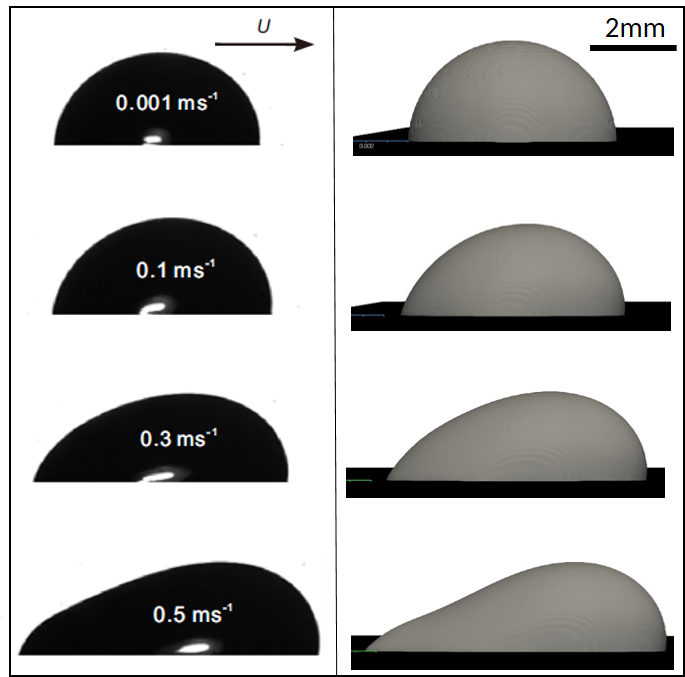}
        \caption{}
        \label{fig-PS-shape}
    \end{subfigure}

    % ===== 整体标题 =====
    \caption{Comparison of experimental and simulation results of a water droplet on a PS-Gold surface. (a) Contact line velocity evolution; (b) droplet aspect ratio (length-to-width: $L_d/W_d$); (c) droplet profile at different velocities
    $U_{cl}$: experimental images (left), simulation results (right).}
    \label{fig:example1}
\end{figure}

Here, we qualitatively and quantitatively compare the simulation results for three different surfaces with the measurements of Ref. \citep{li2023kinetic}.
Fig.~\ref{fig:example1}
displays the results for the
PS-gold surface.
In subfigure \ref{fig-PS-ucl}
it can be seen, that
the simulations successfully capture the main trend of the average contact line velocity $U_{cl}$ to increase with time. From the behavior at
different inclination angles ($\alpha$=25$^\circ$, 50$^\circ$, 60$^\circ$, and 70$^\circ$) it follows, that
the movement is faster
at larger inclination angles
due to an enhancing gravitational acceleration. 
In Figure \ref{fig-PS-LW},
%\textcolor{red}{(Yifan, the comparison with the experiment does not show in detail the experimental results for the different inclination angles. Can this somehow be improved?)}
the aspect ratio $L_d/W_d$ increases with $U_{cl}$, indicating that the droplet elongates as it accelerates. The simulation closely follows the experimental trend
and show only minor differences between the cases of different inclination angles.
Hoewever, 
the $L_d/W_d$ values
obtained in the simulations
are systematically 5-10\% lower than in the measurements.
This minor discrepancy may be caused by unresolved micro-scale 
effects of the
contact line dynamics
in the simulations and potential surface roughness effects in experiments that are not modeled.
Figure~\ref{fig-PS-shape} shows good agreement between simulated and measured droplet profiles over a range of contact line velocities, supporting the model’s predictive capability. 
Because the deformation depends primarily on the contact line velocity, profiles were compared at matched contact line velocities $U_{cl}$ rather than at identical inclined angles: for each inclination, the simulation snapshot with the $U_{cl}$ 
attained in the experiment was selected and coompared with the experimental contour.
%Moreover, the droplet shape comparison in Figure \ref{fig-PS-shape} demonstrates good agreement between simulated and observed profile at various velocities, further validating the model’s predictive ability.

\begin{figure}[htbp]
    \centering
    % ===== 左侧 minipage，包含 (a) 和 (b) 两个子图 =====
    \begin{minipage}[c]{0.38\textwidth}
        % -- 子图 (a) --
        \begin{subfigure}[c]{\textwidth}
            \centering
            \includegraphics[width=\textwidth]{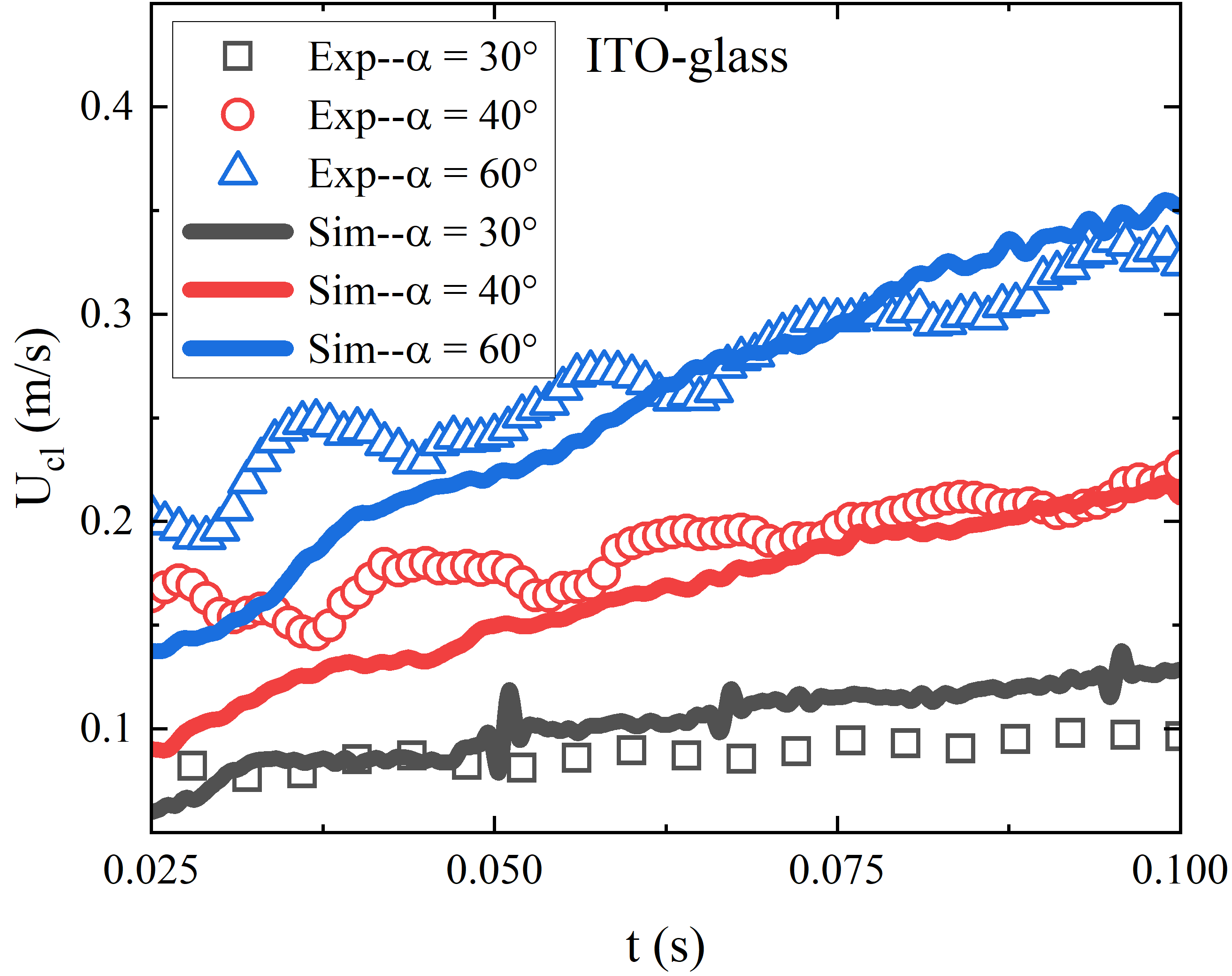}
            \caption{}
            \label{fig-ITO-ucl}
        \end{subfigure}
        \vspace{0.5cm}
        % -- 子图 (b) --
        \begin{subfigure}[c]{\textwidth}
            \centering
            \includegraphics[width=0.96\textwidth]{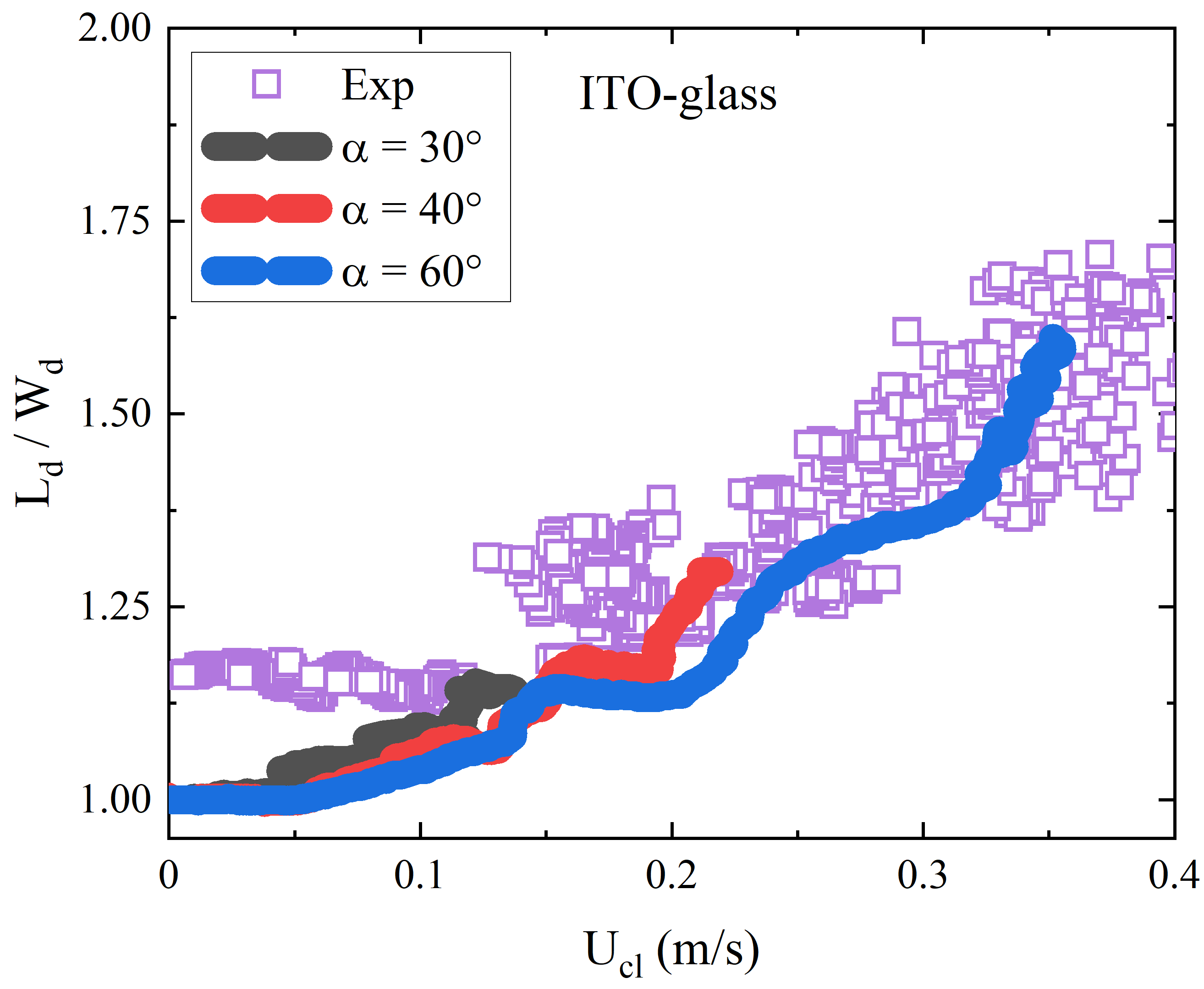}
            \caption{}
            \label{fig-ITO-LW}
        \end{subfigure}
    \end{minipage}
    \hfill
    % ===== 右侧子图 (c) =====
    \begin{subfigure}[c]{0.6\textwidth}
        \centering
        \includegraphics[width=\textwidth]{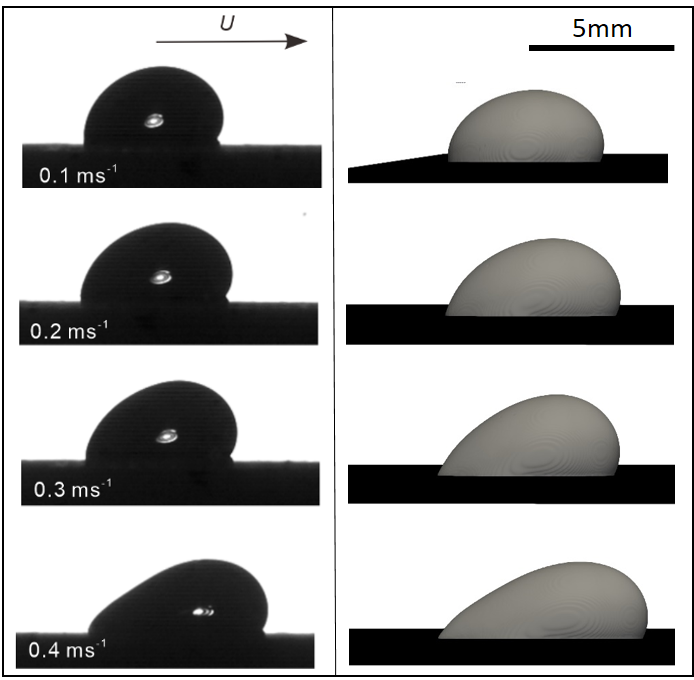}
        \caption{}
        \label{fig-ITO-shape}
    \end{subfigure}

    % ===== 整体标题 =====
    \caption{Comparison of experimental and simulation results of a water droplet sliding on an ITO-glass surface. (a) Contact line velocity evolution; (b) velocity-dependent aspect ratio (length-to-width: $L_d/W_d$); (c) droplet profile at different velocities
    $U_{cl}$
    : experimental images (left), simulation results (right).}
    \label{fig:example2}
\end{figure}

Similarly, for the ITO-glass surface, the temporal evolution of the 
contact line velocity shown in Figure \ref{fig-ITO-ucl} agrees qualitatively
well with experimental results across three different inclination angles. 
While some alternating behavior 
of acceleration and deceleration is
observed in the experimental data, the overall trend and magnitude are 
approximately captured by the simulations. These show a mainly monotonic increase of the sliding
velocity, corresponding to a slightly
larger main acceleration than in the
measurements. These differences might
be caused by unconsidered surface roughness and also by smaller droplet sizes used in the experiments (see above).
The larger deviation at early times is due to differences in the initial
conditions of experiments and simulations. As mentioned above,
the simulations start from an ideal spherical cap with a relaxed contact line, whereas the experiments begin after the feeding needle is withdrawn.
Here, the droplet may already have
started to deform and move on the surface, which explains the 
%which induces a transient acceleration and yields a 
higher initial contact line velocity $U_{cl}$.
Figure \ref{fig-ITO-LW} shows 
the behavior of
the droplet aspect ratio versus
$U_{cl}$. The simulation results follow the general trend of the aspect ratio
to grow with velocity, with the overlay of step-like behavior as discussed above.
%that the droplet aspect ratio increases consistently with $U_{cl}$, and the simulation data align with the experimental observations. 
However, for small contact line velocities $U_{cl}<$ 0.1 m/s,
the aspect ratio measured is initially
by about 20\%  larger than unity, as  expected for the simulations.
This is due to the differences in the
initial conditions of experiments and
simulations mentioned above.
As can be seen,
this gap diminishes considerably as the sliding velocity increases.
In Figure \ref{fig-ITO-shape}, the comparison of droplet profiles at four representative velocities again shows a strong match between experimental images and simulation snapshots, confirming the robustness of our numerical approach in predicting the complex dynamics of sliding droplets on different surfaces.

\begin{figure}[htbp]
    \centering
    % ===== 左侧 minipage，包含 (a) 和 (b) 两个子图 =====
    \begin{minipage}[c]{0.38\textwidth}
        % -- 子图 (a) --
        \begin{subfigure}[c]{\textwidth}
            \centering
            \includegraphics[width=\textwidth]{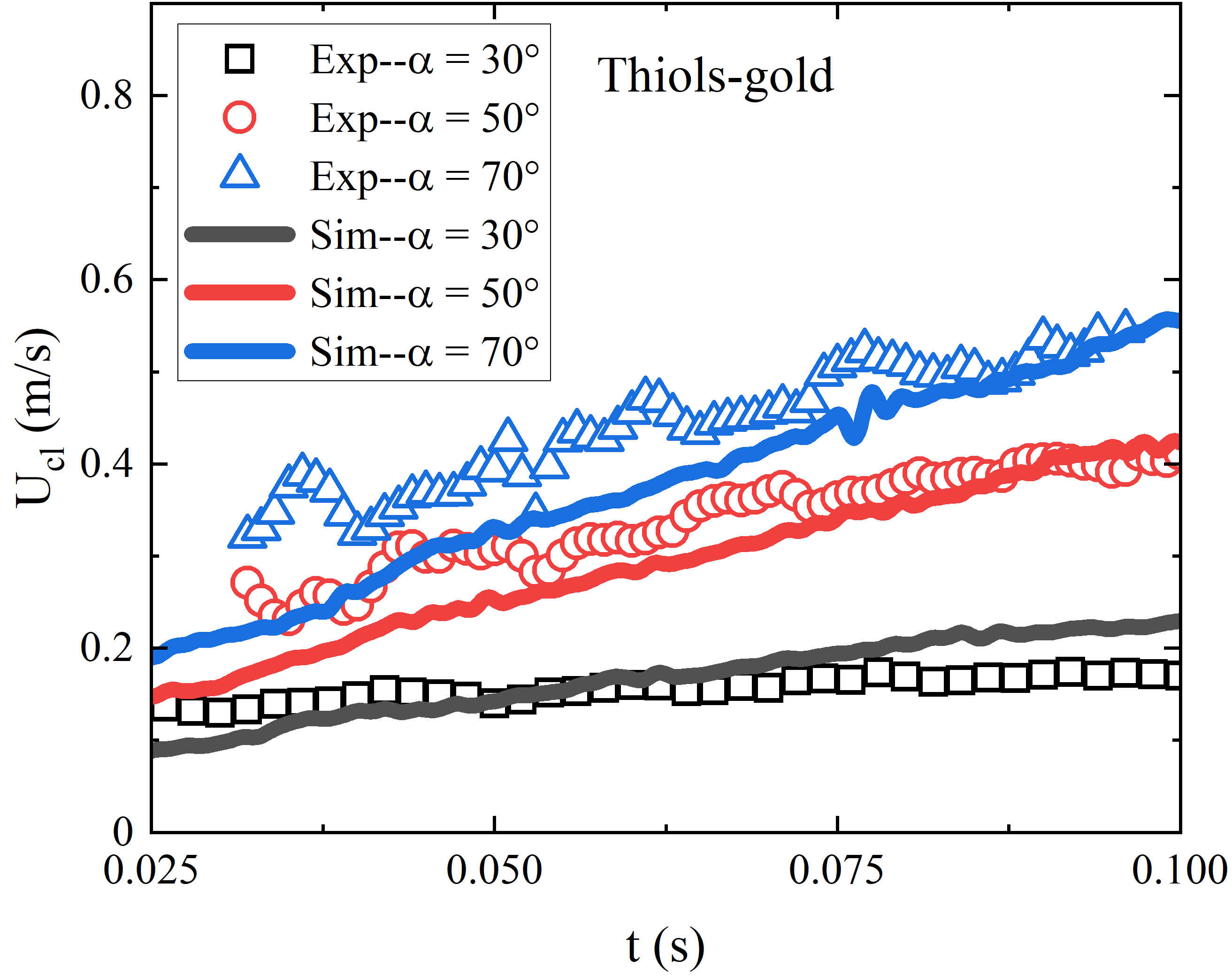}
            \caption{}
            \label{fig-Thiols-ucl}
        \end{subfigure}
        \vspace{0.5cm}
        % -- 子图 (b) --
        \begin{subfigure}[c]{\textwidth}
            \centering
            \includegraphics[width=0.96\textwidth]{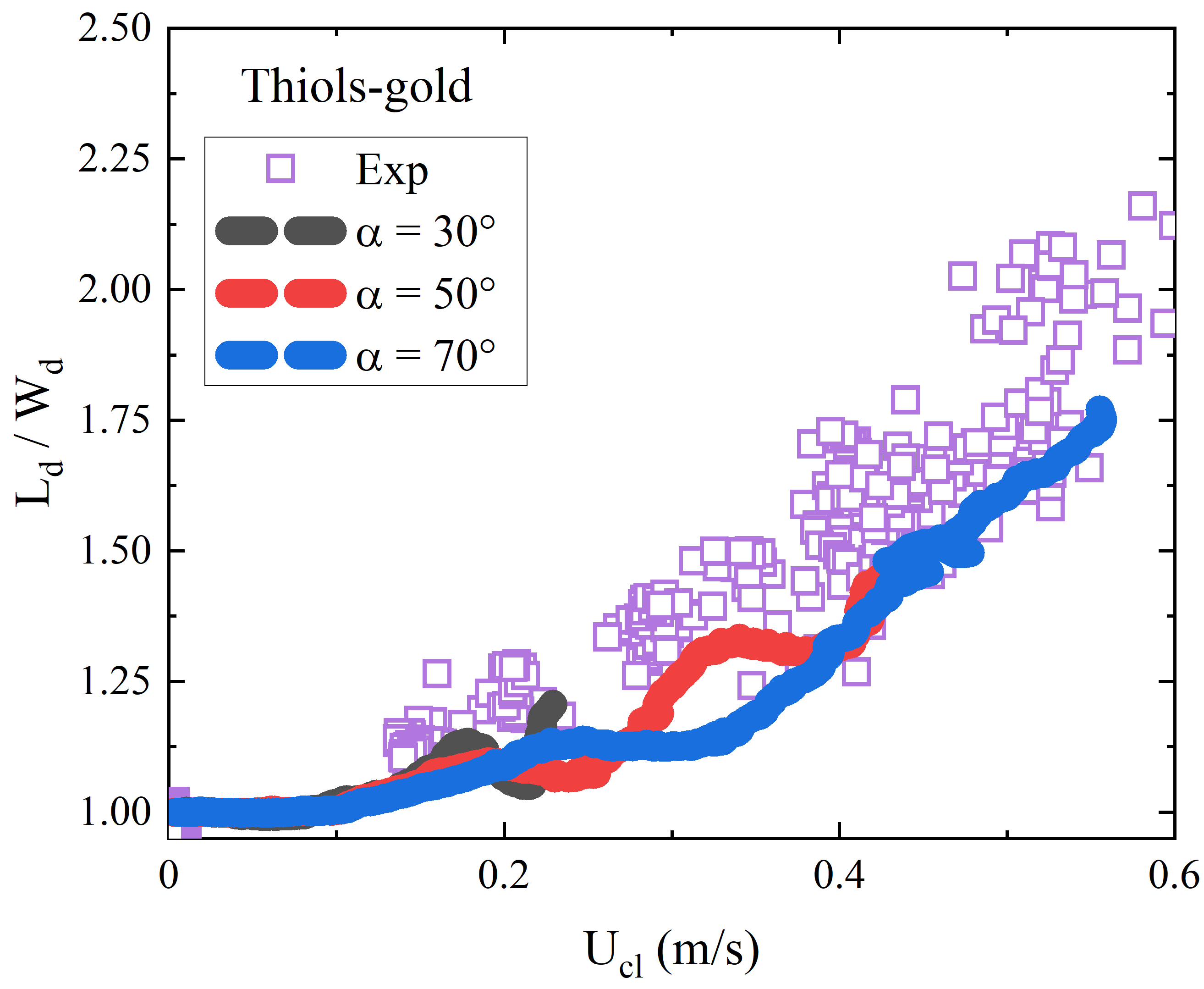}
            \caption{}
            \label{fig-Thiols-LW}
        \end{subfigure}
    \end{minipage}
    \hfill
    % ===== 右侧子图 (c) =====
    \begin{subfigure}[c]{0.6\textwidth}
        \centering
        \includegraphics[width=\textwidth]{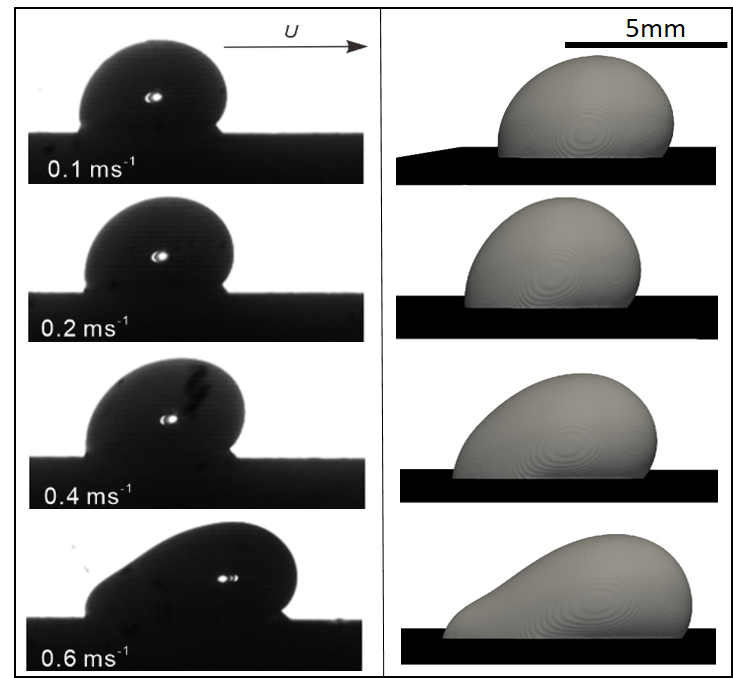}
        \caption{}
        \label{fig-Thiols-shape}
    \end{subfigure}

    % ===== 整体标题 =====
    \caption{Comparison of experimental and simulation results of a water droplet on a Thiols surface. (a) Contact line velocity evolution; (b) velocity-dependent aspect ratio (length-to-width: $L_d/W_d$); (c) droplet profile at different velocities
    $U_{cl}$: experimental images (left), simulation snapshots (right).}
    \label{fig:example3}
\end{figure}

Figure \ref{fig-Thiols-ucl} shows that 
for
the Thiols surface
also a qualitatively
good agreement can be stated
between the simulated and measured contact line velocities 
at different inclination angles. 
The hydrophobicity of the Thiols surface ($\theta_s$ = 120$^\circ$) yields characteristically higher $U_{cl}$ values compared to more wetted surfaces, and the simulation reproduces both the velocity magnitude and acceleration profile. %\textcolor{red}{(please again comment on the initially large differences.)}
The simulations again show a slightly smoother 
behavior of the contact line velocity
compared to the experiments,
which exhibit more fluctuations, likely to be caused by surface heterogeneities. Besides, the acceleration is found to be slightly larger in the simulations, for the
reasons already discussed above.
%or experimental noise.
As seen in Figure \ref{fig-Thiols-LW}, for the aspect ratio evolution 
both, simulations and measurements,
follow the expected linear 
relationship with $U_{cl}$, 
whereby the overlayed
step-like behavior discussed above
is only visible in the simulations.
The aspect ratio is found to be slightly larger in the measurements,
which might be caused by additional
pinning due to surface inhomogeneities.
A close correspondence 
is visible also in
the droplet shapes
shown in Figure \ref{fig-Thiols-shape}, where remarkable agreement between simulated and experimental results
becomes visible
at four characteristic velocities.
At higher velocities, the droplet adopts an asymmetric teardrop morphology, characterized by a flattened advancing contact line and an elongated, elevated receding edge, 
a distinctive deformation pattern induced by the hydrophobicity of the substrate.
The combined results validate the model's capability to handle increased contact angle hysteresis and discontinuous motion regimes typical of hydrophobic surfaces.

\subsection{Dynamic versus constant contact angle}
\label{dca-vs-cst-models}

In order to gain insight into the role of dynamic wetting
in droplet sliding,
in this subsection,
we do a comparison between simulations
with dynamic wetting
and with a constant
contact angle applied.
We consider droplet sliding on an PS-gold surface with an
inclination angle of
$\alpha = 50^\circ$. 
The dynamic contact angle (DCA) model 
used is based on the
fitted wetting curve
shown in 
Fig.~\ref{fig-ps-gold}, and the constant contact angle (Cst CA) applied is the corresponding
static contact angle
of $\theta_s = 88^\circ$.
%\st{The comparison highlights the critical role of dynamic wetting behavior in accurately reproducing experimental observations of droplet motion and deformation.}

\begin{figure}[htbp]
    \centering
    %%%%% 第一行 (a) (b) %%%%%
    \begin{subfigure}[b]{0.48\textwidth}
        \centering
        \includegraphics[width=\textwidth]{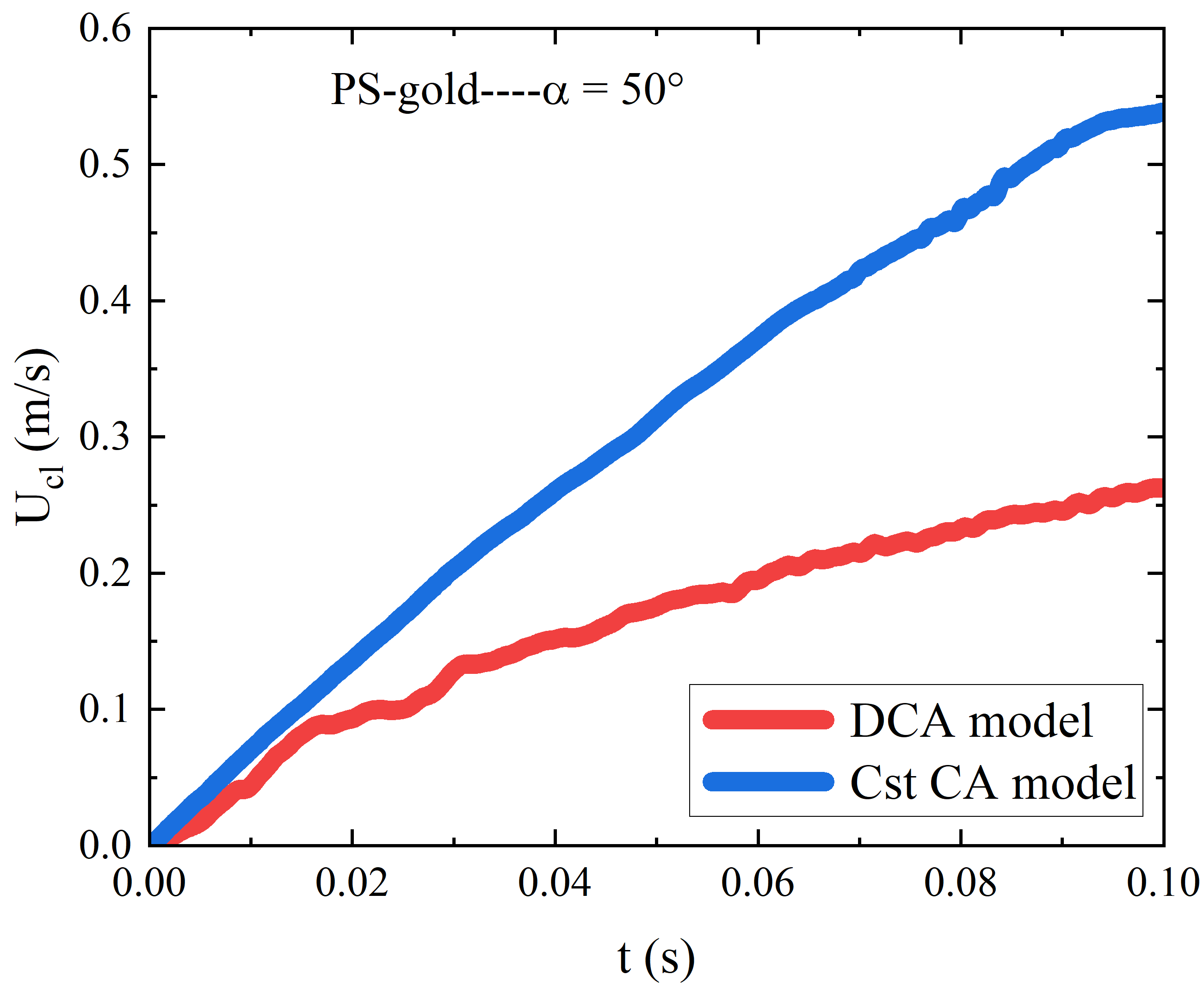}
        \caption{}
        \label{fig-dca-ca-ucl}
    \end{subfigure}
    \hfill
    \begin{subfigure}[b]{0.46\textwidth}
        \centering
        \includegraphics[width=\textwidth]{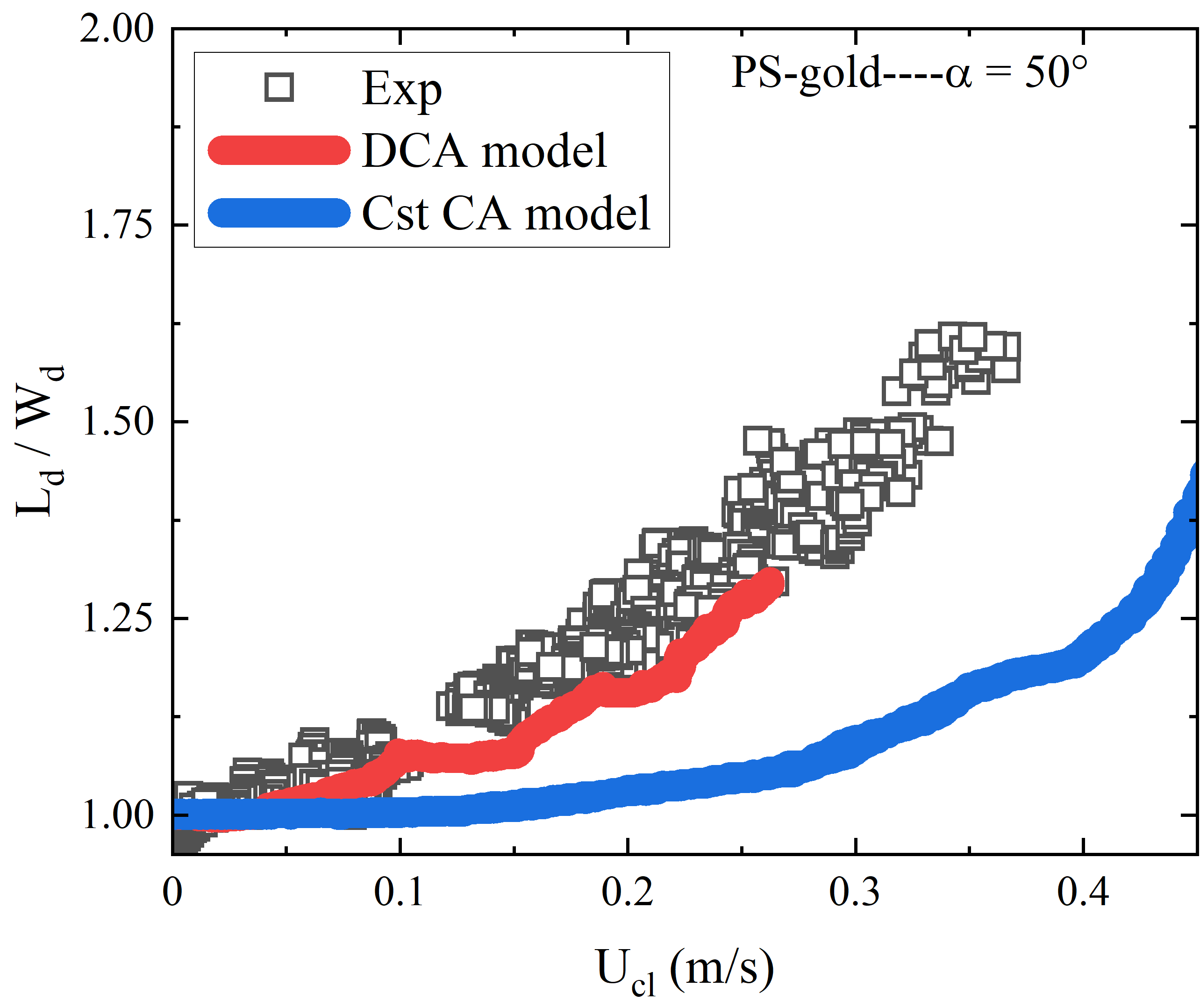}
        \caption{}
        \label{fig-dac-ca-LW}
    \end{subfigure}
    
    \vspace{1em} % 可根据需要调整行间距
    
    %%%%% 第二行 (c) (d) %%%%%
    \begin{subfigure}[b]{0.52\textwidth}
        \centering
        \includegraphics[width=\textwidth]{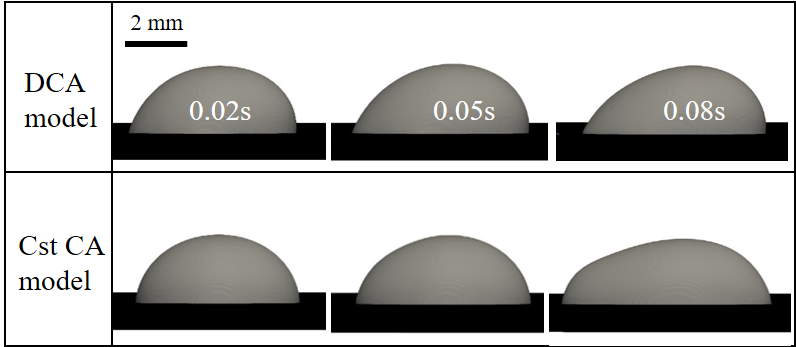}
        \caption{}
        \label{fig-dac-side-view}
    \end{subfigure}
    \hfill
    \begin{subfigure}[b]{0.44\textwidth}
        \centering
        \includegraphics[width=\textwidth]{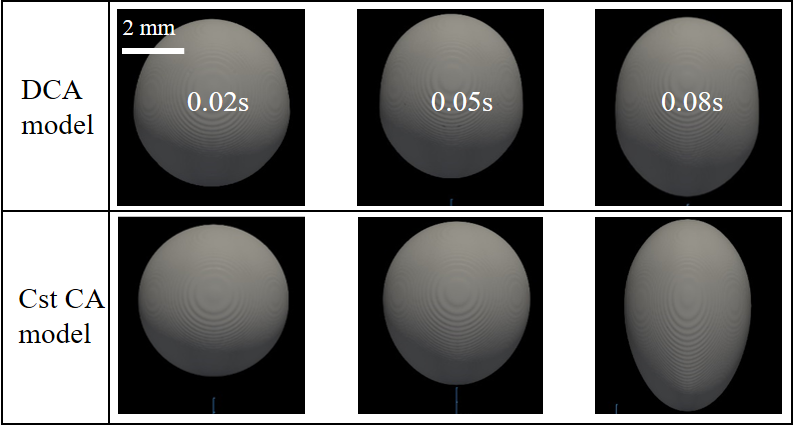}
        \caption{}
        \label{fig-dac-top-view}
    \end{subfigure}
    
    \caption{
        Comparison of dynamic and constant contact angle models in
        simulations of
        droplet sliding 
        at $\alpha=50^\circ$ on a PS-gold surface. 
        (a) Evolution
        of the
        contact line velocity. 
        (b) velocity-dependent aspect ratio ($L_d/W_d$).
        The squares denote experimental data from
      \citep{li2023kinetic}.
        (c) and (d): side and top views of the
        droplet shape
        at different time instants.
    }
    \label{fig:4-2x2}
\end{figure}

The results of both
simulations are shown in Figure 
\ref{fig:4-2x2}.
In Figure \ref{fig-dca-ca-ucl}, the evolution of contact line velocity $U_{cl}$ over time reveals that the DCA model produces a more gradual and physically realistic acceleration of the droplet. 
In contrast, the Cst CA model significantly overestimates the velocity across the entire time range, leading to an accelerated droplet motion.
Figure \ref{fig-dac-ca-LW} shows the dependence of the droplet aspect ratio $L_d/W_d$ on $U_{cl}$. 
The DCA model tracks the experimental measurements well, reproducing the moderate increase in elongation as the droplet accelerates. 
On the other hand, the Cst CA model fails to reproduce the trend observed in both experimental data and DCA model.
Even at high velocities $U_{cl}>$ 0.3 m/s, the Cst CA model underestimates the droplet elongation and yields
aspect ratios 
smaller than 1.5,
whereas the measured
values are clearly larger than 1.5.
This deviation arises because a fixed contact angle prevents asymmetric contact line motion. 
Both the front and back edges of the droplet move similarly, suppressing the differential advancing and receding required to generate a realistic elongated shape.
The side and top view comparisons of the
droplet shapes
in Figure \ref{fig-dac-side-view} and \ref{fig-dac-top-view} further illustrate the differences. 
As seen before, the DCA model yields droplet shapes that closely resemble realistic experimental behavior, characterized by   elongated and stable
profiles with progressive spreading. In contrast, the Cst CA model results in overly flattened and elongated droplets, particularly evident in the top views. 
Therefore, it is essential to incorporate a dynamic contact angle model 
into numerical simulations
in order to 
capture the true kinematics and deformation of
droplets sliding on inclined surfaces
due to gravitational acceleration.
%--------------------------------------------
%--------------------------------------------

\section{Conclusion}
\label{conclusion}
This study presents the implementation of a three-dimensional dynamic wetting model within the Basilisk framework, grounded in a geometric Volume-of-Fluid (VOF) formulation. 
A key innovation is the introduction of a geometric interpolation scheme for contact line velocity, which enhances both numerical accuracy and stability, particularly in the vicinity of the moving contact line.
The model further incorporates a physically motivated dynamic contact angle formulation, including contact angle hysteresis (CAH), enabling a realistic representation of advancing and receding behaviors.

Extensive validation against experimental data—including droplet spreading, splashing, and sliding on various substrates—demonstrates that the model achieves quantitative agreement with the measured droplet dynamics. 
The proposed methodology successfully preserves axisymmetry in inherent configurations, as confirmed by 3D simulations reproducing axisymmetric benchmarks of droplet spreading. 
This consistency highlights the robustness of the 3D framework in extending beyond conventional 2D axisymmetric solvers.
Moreover, the comparison between dynamic and static contact angle models for a sliding droplet underscores the necessity of capturing dynamic wetting phenomena to accurately reproduce observed motion. 
The dynamic model shows a marked improvement in predictive capability, especially in regimes dominated by contact line motion.

In summary, this work provides a comprehensive and extensible numerical framework for dynamic wetting in three dimensions, offering enhanced accuracy, stability, and versatility for simulating droplet and bubble interactions in diverse physical settings.
Beyond droplet-scale phenomena, the framework offers potential for simulating a wide range of interfacial processes, such as bubble coalescence, detachment under shear flow. 
Its ability to maintain accuracy under complex boundary conditions and large deformations highlights its utility in both fundamental and applied multiphase flow studies.

%===================================%
\section{Acknowledgement}
%===================================%
We thank Gabriele Gennari 
and Mengyuan Huang
for fruitful discussions.
We are very grateful to the open-minded
community maintaining and developing
the freely available code framework Basilisk (http://basilisk.fr). We will make our development available there.
We gratefully acknowledge financial support by the Federal Ministry of Education and Research (BMBF)
within the project H2Giga-SINEWAVE, grant no.~03HY123E.
This work was further supported by the German Space Agency (DLR) with funds provided by the Federal
Ministry of Economics and Technology (BMWi) due to an enactment of the German Bundestag under grant
no. DLR 50WM2058 (project MADAGAS II).

%\appendix
%\section{Appendix}

%\bibliographystyle{alpha}
%\bibliographystyle{elsarticle-harv}  
%\biboptions{authoryear,round}      
\bibliography{sample}

\end{document}